\documentclass[12pt]{article}
\usepackage[centertags]{amsmath}
\usepackage{amsfonts}
\usepackage{amssymb}
\usepackage{amsthm}
\usepackage{newlfont}

\newtheorem{theorem}{Theorem}
\newtheorem{lemma}{Lemma}

\newtheorem{corollary}{Corollary}

\begin{document}

\title{On one-dimensional G-dynamics and non-Hermitian Hamiltonian operators}
\author{Jack Whongius\thanks{School of Mathematical Sciences,~ Xiamen
University,~ 361005,  China.~ \newline The research of the author is  partially  supported by NSF grants.\newline The author would like to thank Professor Lee for the valuable suggestions and some discussions in designing this paper.\newline Email:  fmsswangius@stu.xmu.edu.cn. }}
 \maketitle

\par  Focusing on the algebraical analysis of two various kinds of one-dimensional G-dynamics  ${{\hat{w}}^{\left( cl \right)}}$ and ${{\hat{w}}^{\left( ri\right)}}$ separately induced by different Hamiltonian operators $\hat{H} $ are the keypoints.  In this work, it's evidently proved that an identity ${{\hat{w}}^{\left( cl \right)}}{{u}^{-1/2}}\equiv0$ always holds for any $u>0$ based on the formula of one-dimensional G-dynamics ${{\hat{w}}^{\left( cl \right)}}$. We prove that the G-dynamics ${{\hat{w}}^{\left( cl \right)}}$ and ${{\hat{w}}^{\left( ri\right)}}$ obey Leibniz identity if and only if ${{\hat{w}}^{\left( cl \right)}}1=0$ and ${{\hat{w}}^{\left( ri\right)}}1=0$, respectively. \par
In accordance with the G-dynamics ${{\hat{w}}^{\left( cl \right)}}$,   we investigate the unique eigenvalues equation ${{\hat{w}}^{\left( cl \right)}}L\left( u,t,\lambda  \right)=-\sqrt{-1}\lambda L\left( u,t,\lambda  \right)$ of the G-dynamics with a precise geometric eigenfunction $L\left( u,t,\lambda  \right)={{u}^{-1/2}}{{e}^{{{\lambda }}t}},~u>0$ as time $t\in \left[ 0,T \right]$ develops and the equation of energy spectrum is then induced. The non-Hermitian Hamiltonian operators are studied as well, we obtain a series of ODE with their special solutions, and we prove multiplicative property of the geometric eigenfunction. The coordinate derivative and time evolution of the G-dynamics are respectively considered.
Seeking the invariance of G-dynamics ${{\hat{w}}^{\left( cl \right)}}$  under coordinate transformation is considered, so that we think of one-dimensional G-dynamics ${{\hat{w}}^{\left( cl \right)}}$ on coordinate transformation, it gives the rule of conversion between two coordinate systems. As a application, some examples are given for such rule of conversion. Meanwhile, we search the conditions that quantum geometric bracket vanishes and a specific case follows.


\section{Introduction of the G-dynamics}
\vspace{.2 cm}
\ \ \ \ \ \
The G-dynamics, introduced by Wang [GW, \S 7.1] in 2020s, is the quantum
evolution equation
 $\hat{w}=-\sqrt{-1}{{\left[ s,\hat{H} \right]}_{QPB}}/\hbar$
for a geometric function $s$ determined by the manifolds.
The quantum Poisson brackets $\left[ \hat{f},\hat{g} \right]_{QPB}=\hat{f}\hat{g}-\hat{g}\hat{f}$ is the commutator of two operators $\hat{f},~\hat{g}$, in the Heisenberg picture (1925s) [DG], the operators evolve with time described by Heisenberg equation defined by quantum Poisson brackets and the wave functions remain constant. Later, Wang [\S 7] has generally proposed the concepts of the quantum covariant Hamiltonian system $\left[ \hat{f},\hat{H} \right]$ defined by the quantum covariant Poisson bracket $\left[ \hat{f},\hat{g} \right]={{\left[ \hat{f},\hat{g} \right]}_{QPB}}+G\left( s,\hat{f},\hat{g} \right)$ with the complete additional item quantum geometric bracket $G\left(s,\hat{f},\hat{g} \right)=\hat{f}{{\left[ s,\hat{g} \right]}_{QPB}}-\hat{g}{{\left[ s,\hat{f} \right]}_{QPB}}$ for two operators $\hat{f},~\hat{g}$, introducing a geometric real-valued smooth scalar function as a structural function or the potential function $s\in {{C}^{\infty}}\left( M \right)$ is completely determined by the corresponding manifold space.
As a result of the quantum covariant Hamiltonian system, it forms the covariant dynamics including the G-dynamics. In another aspect, if quantum geometric bracket disappears, that is to say, $G\left(s,\hat{f},\hat{g} \right)=0$ holds for operators $\hat{f},~\hat{g}$, then the quantum covariant Poisson bracket is degenerated to the quantum Poisson bracket, as such case, it gets an equality shown by
$\hat{f}{{\left[ s,\hat{g} \right]}_{QPB}}=\hat{g}{{\left[ s,\hat{f} \right]}_{QPB}}$. Above all, with such interesting characters, and particular features of the quantum covariant Poisson bracket and quantum geometric bracket,
such motivation has provided a new mathematical tool for us to reconsider what we see the quantum mechanics.

The most general form of Schr\"{o}dinger equation is the time-dependent equation (1926s) as a linear partial differential equation, which describes a system evolving with time
\begin{equation}\label{a4}
  \sqrt{-1}\hbar {{\partial }_{t}}\varphi ={{\hat{H}}^{\left( cl \right)}}\varphi
\end{equation}
where $\partial/\partial t$ symbolizes a partial derivative in terms of time $t$, $\varphi$  is the wave function of the quantum system.
\begin{equation}\label{b5}
  {{\hat{H}}^{\left( cl \right)}}=-\frac{{{\hbar }^{2}}}{2m}{d}^{2}/d{{x}^{2}}+V\left( x \right)
\end{equation}
is the one dimensional Hamiltonian operator in terms of the coordinates, where $V$ is the potential energy in terms of the position $x$.  Note that the one-dimensional G-dynamics ${{\hat{w}}^{\left( cl \right)}}=-\sqrt{-1}{{\left[ s,{{\hat{H}}^{\left( cl \right)}} \right]}_{QPB}}/\hbar$ is precisely induced by the classical Hamiltonian operator ${{\hat{H}}^{\left( cl \right)}}$ \eqref{a4}.
The geometric wave equation induced by the G-dynamics is given by
\begin{equation}\label{e3}
  \sqrt{-1}\hbar \hat{w}\psi ={{\hat{H}}^{\left( \operatorname{Im} \right)}}\left( \hat{w} \right)\psi
\end{equation}
for a given wave function $\psi$, where ${{\hat{H}}^{\left( \operatorname{Im} \right)}}\left( \hat{w} \right)={{\left[ s,\hat{H} \right]}_{QPB}}$ is the geometric Hamiltonian operator.
Obviously, this equation \eqref{e3} is highly similar to the
Schr\"{o}dinger equation \eqref{a4} in form.
The Schr\"{o}dinger equation and the geometric wave equation have the analogous expression.

For instance, to organize two equations in one dimensional form together is given as follows:
\begin{enumerate}
  \item ~ Schr\"{o}dinger equation: $\sqrt{-1}\hbar {{\partial }_{t}}\varphi ={{\hat{H}}^{\left( cl \right)}}\varphi$,
      where ${{\hat{H}}^{\left( cl \right)}}=-\frac{{{\hbar }^{2}}}{2m}\frac{{{d}^{2}}}{d{{x}^{2}}}+V\left( x \right)$ is the classical Hamiltonian operator  \eqref{b5}.
  \item ~ Geometric wave equation: $\sqrt{-1}\hbar {{\hat{w}}^{\left( cl \right)}}\psi ={{\hat{H}}^{\left( \operatorname{Im} \right)}}({\hat{w}}^{\left( cl \right)})\psi$,
    geometric Hamiltonian operator ${{\hat{H}}^{\left( \operatorname{Im} \right)}}({\hat{w}}^{\left( cl \right)})={{\left[ s,{{\hat{H}}^{\left( cl \right)}} \right]}_{QPB}}=\frac{{{\hbar }^{2}}}{2m}\left( 2u\frac{d}{dx}+\frac{du}{dx} \right)$ is closely corresponding to the classical Hamiltonian operator ${{\hat{H}}^{\left( cl \right)}}$ \eqref{b5} given by the Schr\"{o}dinger equation  \eqref{a4}.
\end{enumerate}
Note that the Schr\"{o}dinger equation is a second order differential equation.
Obviously, these two equations above are parallel to each other and independent to each other,  the geometric wave equation can be to get the solution.  By sorting out the knowledge system as above displays, surely, the geometric wave equation \eqref{e3} has a special status in quantum mechanics.  Due to the solution of the geometric wave equation \eqref{e3}  can be directly obtained, and meanwhile, the solution shows us a diverse form different from what we see in quantum mechanics,   then it has necessary to analyze why it exactly means, geometrically.~ In order to understand G-dynamics ${{\hat{w}}^{\left( cl \right)}}$ better, we give a complete deduction of the one-dimensional G-dynamics ${{\hat{w}}^{\left( cl \right)}}$ as follows
\begin{align}\label{b6}
{{{\hat{w}}}^{\left( cl \right)}}\psi &=\frac{1}{\sqrt{-1}\hbar }{{\left[ s,{{\hat{H}}^{\left( cl \right)}} \right]}_{QPB}}\psi =\frac{1}{\sqrt{-1}\hbar }{{\left[ s,-\frac{{{\hbar }^{2}}}{2m}\frac{{{d}^{2}}}{d{{x}^{2}}}+V\left( x \right) \right]}_{QPB}}\psi  \notag\\
 & =\frac{\sqrt{-1}\hbar }{2m}{{\left[ s,\frac{{{d}^{2}}}{d{{x}^{2}}} \right]}_{QPB}}\psi =\frac{\sqrt{-1}\hbar }{2m}\left( s\frac{{{d}^{2}}\psi }{d{{x}^{2}}}-\frac{{{d}^{2}}}{d{{x}^{2}}}\left( \psi s \right) \right) \\
 & =\frac{\sqrt{-1}\hbar }{2m}\left( s{{\psi }_{xx}}-\psi {{s}_{xx}}-2{{s}_{x}}{{\psi }_{x}}-s{{\psi }_{xx}} \right) \notag\\
 & =-\frac{\sqrt{-1}\hbar }{2m}\left( \psi {{s}_{xx}}+2{{s}_{x}}{{\psi }_{x}} \right) \notag\\
 & ={{b}_{c}}\left( 2u{{\psi }_{x}}+\psi {{u}_{x}} \right) \notag
\end{align}for a given wave function $\psi$ in a clear expression,
where $u={{s}_{x}}=ds/dx,{{u}_{x}}={{s}_{xx}}$, $b_{c}=-\sqrt{-1}\hbar/2m$ and $\frac{{{d}^{2}}}{d{{x}^{2}}}\left( \psi s \right)=\psi {{s}_{xx}}+2{{s}_{x}}{{\psi }_{x}}+s{{\psi }_{xx}}$, and  ${{\left[ s,V\left( x \right) \right]}_{QPB}}=0$,  then it leads us to the result ${{\hat{w}}^{\left( cl \right)}}={{b}_{c}}\left( 2ud/dx+{{u}_{x}} \right)$.

The G-dynamics in one-dimensional form, as mentioned, ${{\hat{w}}^{\left( cl \right)}}$ is expressed as
\begin{equation}\label{e1}
{{\hat{w}}^{\left( cl \right)}}={{b}_{c}}\left( 2u\frac{d}{dx}+\frac{du}{dx} \right)={{b}_{c}}\hat{Q}
\end{equation}where line curvature $u=ds/dx$ has been used, $s$ is a scalar valued function on $M$.
The G-dynamics is quasi-one dimensional. As this formula \eqref{e1} expressed, the main part is the curvature operator expressed as
\begin{equation}\label{a8}
  \hat{Q}={{\hat{w}}^{\left( cl \right)}}/{{b}_{c}}=2u\frac{d}{dx}+u_{x}=2m{{\left[ s,{{\hat{H}}^{\left( cl \right)}} \right]}_{QPB}}/{{\hbar }^{2}}
\end{equation}where $u_{x}=du/dx$ is denoted, and
the left side means the pure geometric operator while the right side stands for the physics.
This means the G-dynamics stands for the properties of the space. The G-dynamics links to the geometry, the continuous geometric function $u$ is in charge of the curve of the space.
Obviously, we can easily check out
${{\hat{w}}^{\left( cl \right)}}$ \eqref{e1} to a more compact form by using the one dimensional velocity operator ${{\hat{v}}^{\left( cl \right)}}$,
\begin{equation}\label{b2}
  {{\hat{w}}^{\left( cl \right)}}=u{{\hat{v}}^{\left( cl \right)}}+\frac{1}{2}{{\hat{v}}^{\left( cl \right)}}u
\end{equation}
where ${{\hat{v}}^{\left( cl \right)}}={{\hat{p}}^{\left( cl \right)}}/m=2b_{c}d/dx$ can be realized as the one dimensional velocity operator, and meanwhile, it gets  ${{\left[ s,{{\hat{p}}^{\left( cl \right)}} \right]}_{QPB}}=\sqrt{-1}\hbar u$. Hence, it produces $u=-\sqrt{-1}{{\left[ s,{{\hat{p}}^{\left( cl \right)}} \right]}_{QPB}}/\hbar $. Therefore, we have
\begin{equation}\label{c8}
  {{\left[ s,{{{\hat{p}}}^{\left( cl \right)}} \right]}_{QPB}}=\sqrt{-1}\hbar u,~~{{\left[ s,{{{\hat{H}}}^{\left( cl \right)}} \right]}_{QPB}}=\sqrt{-1}\hbar {{\hat{w}}^{\left( cl \right)}}
\end{equation}
Clearly, this formula \eqref{b2} expresses a better structure to help us understand the G-dynamics ${{\hat{w}}^{\left( cl \right)}}$ \eqref{e1}. Conclusively, we obtain the similar mode for the derivation of the function $u$ and G-dynamics ${{{\hat{w}}}^{\left( cl \right)}}$ as a quantum operator together,
\begin{align}
  & u=-\sqrt{-1}{{\left[ s,{{\hat{p}}^{\left( cl \right)}} \right]}_{QPB}}/\hbar  \notag\\
 & {{{\hat{w}}}^{\left( cl \right)}}=-\sqrt{-1}{{\left[ s,{{\hat{H}}^{\left( cl \right)}} \right]}_{QPB}}/\hbar =u{{\hat{v}}^{\left( cl \right)}}+{{\hat{v}}^{\left( cl \right)}}u/2 \notag
\end{align}Their relation is given by the second formula above.

In this work, we study two kinds of one-dimensional G-dynamics ${{\hat{w}}^{\left( cl \right)}}$ and ${{\hat{w}}^{\left( ri\right)}}$ respectively based on the work of paper [GW]. We start with the fundamental properties of the one-dimensional G-dynamics ${{\hat{w}}^{\left( cl \right)}}$ \eqref{e1}, then we prove an identity ${{\hat{w}}^{\left( cl \right)}}{{u}^{-1/2}}\equiv0$ for any $u>0$ strictly,    consider the eigenvalues equation of the G-dynamics \eqref{e1},  and then we focus on the properties of geometric eigenfunction of the G-dynamics ${{\hat{w}}^{\left( cl \right)}}$ and ${{\hat{w}}^{\left( ri\right)}}$ respectively.

We prove that  there exists a function $\Phi$ such that ${{\hat{w}}^{\left( ri \right)}}\Phi =0$ holds, that is the ODE given by $2u{{\Phi }_{x}}+\Phi {{u}_{x}}+2{{u}^{2}}\Phi =0$,  we find the solution $\Phi(x) ={{C}_{8}}{{u}^{-1/2}}{{e}^{-s}}$, there is an extra terms in contrast to the G-dynamics ${{\hat{w}}^{\left( cl \right)}}$. The coordinate derivative and time evolution of the G-dynamics are respectively considered.  The rule of conversion of one-dimensional G-dynamics ${{\hat{w}}^{\left( cl \right)}}$ on coordinate transformation is obtained with some examples to be verified.
At the same time, we study the conditions that quantum geometric bracket vanishes and we give a specific case to verify such conditions.

\par
\vspace{.4 cm}
\section{Main results and proofs}
In this section, we discuss the properties of the G-dynamics ${{{\hat{w}}}^{\left( cl \right)}}$  as a rotating mechanics to describe the quantum observable in a rotating frequency. In the beginning, we have a theorem given as follows.

\begin{theorem}\label{t2}
For the G-dynamics ${{{\hat{w}}}^{\left( cl \right)}}$ given by \eqref{e1}, it has properties as follows
\begin{equation}\label{a7}
  {{\hat{w}}^{\left( cl \right)}}{{c}_{1}}={{c}_{1}}{{b}_{c}}du/dx
\end{equation}
\begin{equation}\label{a6}
  {{\hat{w}}^{\left( cl \right)}}\left( c_{1}f \right)=c_{1} {{\hat{w}}^{\left( cl \right)}}f
\end{equation}
\begin{equation}\label{a5}
  {{{\hat{w}}}^{\left( cl \right)}}\left( f+g \right)=f{{{\hat{w}}}^{\left( cl \right)}}g+g{{{\hat{w}}}^{\left( cl \right)}}f
\end{equation}
\begin{equation}\label{a2}
  {{{\hat{w}}}^{\left( cl \right)}}\left( fg \right)=f{{{\hat{w}}}^{\left( cl \right)}}g+g{{{\hat{w}}}^{\left( cl \right)}}f-{{b}_{c}}fgdu/dx
\end{equation}
 for functions $f,g$, and constant $c_{1}$.
  \begin{proof}
  Firstly, according to G-dynamics ${{\hat{w}}^{\left( cl \right)}}$ \eqref{e1}, we can see that it's a linear operator in a form of first order, hence, \eqref{a7}, \eqref{a6} and \eqref{a5} hold.
 Secondly, for the \eqref{a2}, it yields
 \begin{align}
 {{{\hat{w}}}^{\left( cl \right)}}\left( fg \right) & ={{b}_{c}}\left( 2u\frac{d\left( fg \right)}{dx}+fg\frac{du}{dx} \right) \notag\\
 & =f{{b}_{c}}\left( 2u\frac{dg}{dx}+g\frac{du}{dx} \right)+g{{b}_{c}}\left( 2u\frac{df}{dx}+f\frac{du}{dx} \right)-{{b}_{c}}fg\frac{du}{dx} \notag\\
 & =f{{{\hat{w}}}^{\left( cl \right)}}g+g{{{\hat{w}}}^{\left( cl \right)}}f-{{b}_{c}}fg\frac{du}{dx} \notag
\end{align}
Therefore, the proof is complete.

  \end{proof}
\end{theorem}
Note that the \eqref{a2} indicates that the G-dynamics ${{{\hat{w}}}^{\left( cl \right)}}$ disobeys the Leibniz rule. In other words, the Leibniz rule does not satisfy any more. Notice that for \eqref{a6}, if we choose $f=1$, then it goes back to \eqref{a7}, meanwhile, it gives a result ${{\hat{w}}^{\left( cl \right)}}1={{b}_{c}}du/dx\neq 0$. The formulae \eqref{a6} and \eqref{a5} together states the G-dynamics ${{{\hat{w}}}^{\left( cl \right)}}$  is a linear operator.
\begin{corollary}
  Based on theorem \ref{t2}, it has
  $${{\hat{w}}^{\left( cl \right)}}{{c}_{1}}={{c}_{1}}{{\hat{w}}^{\left( cl \right)}}1$$
  \begin{equation}\label{b1}
    {{{\hat{w}}}^{\left( cl \right)}}\left( fg \right)=f{{{\hat{w}}}^{\left( cl \right)}}g+g{{{\hat{w}}}^{\left( cl \right)}}f-fg{{\hat{w}}^{\left( cl \right)}}1
  \end{equation}
\end{corollary}
Note that the formula \eqref{b1} does not obey the Leibniz identity any more because of the appearance of the term ${{\hat{w}}^{\left( cl \right)}}1$,

\begin{corollary}
  One-dimensional G-dynamics ${{\hat{w}}^{\left( cl \right)}}$ obeys Leibniz identity if and only if ${{\hat{w}}^{\left( cl \right)}}1=0$.
\end{corollary}
Note that ${{\hat{w}}^{\left( cl \right)}}1=0$ means that $du/dx= 0$, it gives constant line curvature $u=u_{0}$.   
Anyway, the formula \eqref{b1} also can be written as \[\left( {{{\hat{w}}}^{\left( cl \right)}}+{{{\hat{w}}}^{\left( cl \right)}}1 \right)\left( fg \right)=f{{\hat{w}}^{\left( cl \right)}}g+g{{\hat{w}}^{\left( cl \right)}}f\]and it leads to $$\left( {{{\hat{w}}}^{\left( cl \right)}}+{{{\hat{w}}}^{\left( cl \right)}}1 \right)\left( {{f}^{2}} \right)=2f{{\hat{w}}^{\left( cl \right)}}f$$ More importantly,  ${{\hat{w}}^{\left( cl \right)}}$ \eqref{e1} then becomes
\[{{\hat{w}}^{\left( cl \right)}}={{b}_{c}}\hat{Q}=2{{b}_{c}}u\frac{d}{dx}+{{\hat{w}}^{\left( cl \right)}}1\]Thusly, ${{\hat{w}}^{\left( cl \right)}}-{{\hat{w}}^{\left( cl \right)}}1=2{{b}_{c}}ud/dx$ is obtained as well, meanwhile, G-dynamics ${{{\hat{w}}}^{\left( cl \right)}}$ seen from \eqref{b2} can induce $${{\hat{v}}^{\left( cl \right)}}u= 2{{\hat{w}}^{\left( cl \right)}}1,~~ u{{\hat{v}}^{\left( cl \right)}}=2{{b}_{c}}ud/dx$$ without doubts.
As a result, it yields $${{\left[ {{{\hat{v}}}^{\left( cl \right)}},u \right]}_{QPB}}=2{{b}_{c}}\left( {{u}_{x}}-ud/dx \right)$$
For a positive function $f=f(x)$, it always has
\begin{align}
 \frac{{{{\hat{w}}}^{\left( cl \right)}}f}{f} &=2{{b}_{c}}u{{f}_{x}}/f+{{b}_{c}}K=2{{b}_{c}}u{{f}_{x}}/f+{{{\hat{w}}}^{\left( cl \right)}}1 \notag\\
 & ={{b}_{c}}u{{\left( \ln {{f}^{2}} \right)}_{x}}+{{{\hat{w}}}^{\left( cl \right)}}1 \notag
\end{align}Accordingly, it follows that
\[\left| \frac{{{{\hat{w}}}^{\left( cl \right)}}f}{f} \right|=\frac{\left| {{{\hat{w}}}^{\left( cl \right)}}f \right|}{\left| f \right|}=\left| 2{{b}_{c}}u{{f}_{x}}/f+{{{\hat{w}}}^{\left( cl \right)}}1 \right|\le 2\left| {{b}_{c}}u{{f}_{x}}/f \right|+\left| {{{\hat{w}}}^{\left( cl \right)}}1 \right|\]
Therefore, it gives $$\left| {{{\hat{w}}}^{\left( cl \right)}}f \right|\le 2\left| {{b}_{c}} \right|\left| u{{f}_{x}} \right|+\left| {{{\hat{w}}}^{\left( cl \right)}}1 \right|\left| f \right|$$  Thusly, the same thing goes to the imaginary geomenergy that is given by
\[{{\hat{H}}^{\left( cl \right)}}\left( {{{\hat{w}}}^{\left( cl \right)}} \right)f/f=\sqrt{-1}\hbar {{\hat{w}}^{\left( cl \right)}}f/f=2\sqrt{-1}\hbar {{b}_{c}}u{{\left( \ln f \right)}_{x}}+\sqrt{-1}\hbar {{\hat{w}}^{\left( cl \right)}}1\]
Generally, the terms ${{\hat{w}}^{\left( cl \right)}}1\ne 0$ always holds.
To start with an important lemma, it's given as follows.
\begin{lemma}\label{l1}
For ~${{{\hat{w}}}^{\left( cl \right)}}f(x)\equiv0$~ holds if and only if $f(x)\equiv C_{7}{{u}^{-1/2}}$, where $C_{7}$ is a constant.
\begin{proof}
In accordance to the G-dynamics ${{\hat{w}}^{\left( cl \right)}}$ \eqref{e1}, it has equation given by  \[{{b}_{c}}\left( 2u\frac{d}{dx}+du/dx \right)f={{b}_{c}}\left( 2uf_{x}+fu_{x} \right)=2{{b}_{c}}fu d\ln \left( f{{u}^{1/2}} \right)/dx=0\]for function $f\neq 0$, where $du/dx=u_{x}$ is denoted,  then it yields $d\ln \left( f{{u}^{1/2}} \right)=0$, subsequently, it then gets the solution expressed as $f={{C}_{7}}{{u}^{-1/2}}$ for any $u>0$, where $C_{7}$ is a constant.
 On another way, it easily verifies the result below
  \begin{align}\label{b3}
{{{\hat{w}}}^{\left( cl \right)}}{{u}^{-1/2}} & ={{b}_{c}}\left( 2u\frac{d}{dx}+\frac{du}{dx} \right){{u}^{-1/2}}\notag\\
&={{b}_{c}}\left( -{{u}^{-1/2}}\frac{du}{dx}+{{u}^{-1/2}}\frac{du}{dx} \right)\equiv0
\end{align}
where $C_{7}=1$ is chosen above.
\end{proof}
\end{lemma}
Note that lemma \ref{l1} indicates that all ${u}^{-1/2}$ satisfies this identity for all $u>0$, and this solution is unique.  Subsequently,  we prove the existence of the eigenfunction of the G-dynamics  \eqref{e1}, and obtain its certain expression.
Based on the foundation above with the lemma \ref{l1}, we can prove the following theorem,
\begin{theorem}\label{t1}
  The eigenvalues equation of the G-dynamics $\sqrt{-1}{{\hat{w}}^{\left( cl \right)}}$ is given by
  \begin{equation}\label{a1}
    \sqrt{-1}{{\hat{w}}^{\left( cl \right)}}L\left( u,t,\lambda  \right)=\lambda L\left( u,t,\lambda  \right)
  \end{equation}
with the identity
${{\hat{w}}^{\left( cl \right)}}{{u}^{-1/2}}\equiv0$,
where eigenvalues $\lambda$ in terms of geometric eigenfunction $L\left( u,t,\lambda  \right)={{u}^{-1/2}}{{e}^{{{\lambda }}t}}$ are given.
\begin{proof}
It's easy to verify the eigenvalues equation
  \begin{align}
 \sqrt{-1}{{{\hat{w}}}^{\left( cl \right)}}\left( {{u}^{-1/2}}{{e}^{\lambda t}} \right) & =\sqrt{-1}{{b}_{c}}\left( 2u\frac{d}{dx}+\frac{du}{dx} \right)\left( {{u}^{-1/2}}{{e}^{\lambda t}} \right) \notag\\
 & =\sqrt{-1}{{b}_{c}}\left( 2\frac{u}{v}\frac{d\left( {{u}^{-1/2}}{{e}^{\lambda t}} \right)}{dt}+\frac{{{u}^{-1/2}}{{e}^{\lambda t}}}{v}\frac{du}{dt} \right) \notag\\
 & =\sqrt{-1}{{b}_{c}}\left( 2\lambda \frac{u}{v}{{u}^{-1/2}}{{e}^{\lambda t}}-\frac{1}{v}{{u}^{-1/2}}{{e}^{\lambda t}}\frac{du}{dt}+\frac{{{u}^{-1/2}}{{e}^{\lambda t}}}{v}\frac{du}{dt} \right) \notag\\
 & =2\lambda\sqrt{-1}{{b}_{c}}\frac{u}{v}\left( {{u}^{-1/2}}{{e}^{\lambda t}} \right) =\alpha\lambda \left( {{u}^{-1/2}}{{e}^{\lambda t}} \right) \notag
\end{align}
where $2\sqrt{-1}{{b}_{c}}\frac{u}{v}=\alpha =\pm1\in \mathbb{R}$ for the conjugate, and $dx=vdt$ has been used, here we choose $\alpha=1$, then  $$\sqrt{-1}{{{\hat{w}}}^{\left( cl \right)}}\left( {{u}^{-1/2}}{{e}^{\lambda t}} \right) =\lambda \left( {{u}^{-1/2}}{{e}^{\lambda t}} \right)$$
therefore, we complete the proof.

\end{proof}
\end{theorem}
Note that theorem \ref{t1} has given a general formula for the eigenvalues equation and eigenfunction together of the G-dynamics ${{\hat{w}}^{\left( cl \right)}}$, especially, the identity \eqref{b3} always holds for any $u>0$.

In particular, the geometric eigenfunction $L\left( u,t,\lambda  \right)={{u}^{-1/2}}{{e}^{{{\lambda }}t}}$ is unique to describe the wave in such weird way, unlike the wave function we deal with before, it gives us a different quantum visual angle.
Based on the lemma \ref{l1}, we immediately obtain the following corollary.
\begin{corollary}\label{c1}
For a power function ${u}^{-\beta}$ such that   ${{{\hat{w}}}^{\left( cl \right)}}{{u}^{-\beta}}\equiv0$ holds if and only if $\beta=1/2$.
\begin{proof}
Since the formula of G-dynamics ${{\hat{w}}^{\left( cl \right)}}$ given by \eqref{e1}, it yields
 \begin{align}
 {{{\hat{w}}}^{\left( cl \right)}}{{u}^{-\beta}} & ={{b}_{c}}\left( 2u\frac{d{{u}^{-\beta}}}{dx}+{{u}^{-\beta}}\frac{du}{dx} \right)={{b}_{c}}\left( -2\beta{{u}^{-\beta}}\frac{du}{dx}+{{u}^{-\beta}}\frac{du}{dx} \right) \notag\\
 & =\left( -2\beta+1 \right){{u}^{-\beta}}{{{\hat{w}}}^{\left( cl \right)}}1 \notag
\end{align}
where ${{u}^{-\beta}}{{{\hat{w}}}^{\left( cl \right)}}1\neq0$ for all $u>0$, then
  ${{{\hat{w}}}^{\left( cl \right)}}{{u}^{-\beta}}=0$ holds if and only if $ -2\beta+1=0$, hence, $\beta=1/2$ is given. Therefore, we finish the proof.
\end{proof}
\end{corollary}
Conversely, notice that ${{{\hat{w}}}^{\left( cl \right)}}{{u}^{-\beta}}\neq0$ holds for $\beta\neq1/2$, such condition explains that ${{u}^{-\beta}},~\beta\neq1/2$ is the eigenfunction of the G-dynamics ${{\hat{w}}^{\left( cl \right)}}$ that is only
associated with eigenvalues ${{\lambda }^{(u)}}=\left( -2\beta+1 \right){{\hat{w}}^{\left( cl \right)}}1$.

More precisely, for corollary \ref{c1} mentioned above, it means that the eigenfunction ${u}^{-1/2}$ always makes the zero eigenvalues of the G-dynamics ${{\hat{w}}^{\left( cl \right)}}$, it means that the corresponding eigenfunction ${u}^{-1/2}$ is the null space of the G-dynamics ${{\hat{w}}^{\left( cl \right)}}$. In other words, ${{{\hat{w}}}^{\left( cl \right)}}f\equiv0$ holds if and only if $f\equiv C_{7}{{u}^{-1/2}}$. Notice that
\begin{equation}\label{b7}
  {{\hat{w}}^{\left( cl \right)}}{{u}^{-1/2}}=\hat{Q}{{u}^{-1/2}}=0
\end{equation}
always holds together for any $u>0$.

Note that theorem \ref{t1} is widely used for general situation, based on this, we can undoubtedly lead to another precise corollary given by the following corollary.
\begin{corollary}
  The eigenvalues equation of the G-dynamics ${{\hat{w}}^{\left( cl \right)}}$ is given by
\[{{\hat{w}}^{\left( cl \right)}}L\left( u,t,\sqrt{-1}w^{(q)}  \right)=w^{(q)} L\left( u,t,\sqrt{-1}w^{(q)} \right)\]with the identity  \eqref{b3}~ $${{\hat{w}}^{\left( cl \right)}}{{u}^{-1/2}}\equiv0$$~ where eigenvalues $w^{(q)}$ in terms of geometric eigenfunction $L\left( u,t,\sqrt{-1}w^{(q)} \right)={{u}^{-1/2}}{{e}^{{{\sqrt{-1}w^{(q)} }}t}}$ are given.
\begin{proof}
The eigenvalues equation of ${{\hat{w}}^{\left( cl \right)}}$ reads
${{\hat{w}}^{\left( cl \right)}}\psi ={{w}^{\left( q \right)}}\psi$ for the eigenvalues ${w}^{\left( q \right)}$ as a geometric frequency  of the G-dynamics
${{\hat{w}}^{\left( cl \right)}}$,  $\psi$ is the eigenfunction of the G-dynamics ${{\hat{w}}^{\left( cl \right)}}$ that is associated with eigenvalue ${w}^{\left( q \right)}$.

The eigenvalues equation of \eqref{e1} is certainly given by \[{{b}_{c}}\left( 2u\psi_{x}+\psi u_{x} \right)={{w}^{\left( q \right)}}\psi \] where $\psi_{x}=d\psi/dx$.
The geometric wave function $\psi ={{C}_{0}}F \left( u,{{w}^{\left( q \right)}} \right)$ as a unique solution that can be directly computed and given by
\begin{equation}\label{e4}
  F\left( u,{{w}^{\left( q \right)}} \right)={{u}^{-1/2}}{{e}^{ \sqrt{-1}a_{c}^{-1}\int{\frac{{{w}^{\left( q \right)}}}{u}dx} }}={{u}^{-1/2}}g\left( x(t) \right)
\end{equation} where $g\left( t\right)=g\left( x(t) \right)= {{e}^{\eta(t)}}$, and $\eta \left( x\left( t \right) \right)=\sqrt{-1}a_{c}^{-1}\int{\frac{{{w}^{\left( q \right)}}}{u}dx}$ is the phase position and $a_{c}=\hbar /m$, ${{C}_{0}}$ is the constant.

Let's analyze how the part of exponential function \eqref{e4} exactly means
\begin{align}
\eta(t) &=\sqrt{-1}a_{c}^{-1}\int{\frac{{{w}^{\left( q \right)}}}{u}dx}=\sqrt{-1}{{\alpha }^{-1}}\int{\frac{{{w}^{\left( q \right)}}}{v}dx} =\sqrt{-1}{{\alpha }^{-1}}{{w}^{\left( q \right)}}\oint{dt} \notag
\end{align} where line element $dx=vdt$ in one-dimensional case has been used, then it yields
$g\left(t \right)={{e}^{\sqrt{-1}{{\alpha }^{-1}}{{w}^{\left( q \right)}}T}}$, and $\eta(t)=\sqrt{-1}{{\alpha }^{-1}}{{w}^{\left( q \right)}}T$,
hence, it deduces $a_{c}^{-1}\int{\frac{{{w}^{\left( q \right)}}}{u}dx} =\alpha ^{-1} {{w}^{\left( q \right)}}T$, and  $\alpha=\pm1$ is a dimensionless constant representing the conjugate and $\oint{dt}={{T}}=t-0$ is the period with respect to the frequency eigenvalues ${w}^{\left( q \right)}$, then $g\left( x \right)={{e}^{\sqrt{-1}{{\alpha }^{-1}}{{w}^{\left( q \right)}}t}}$.  As a result, the main contribution of the geometric wave function \eqref{e4} is rewritten and simplified in a precise form
$$F \left( u,t,{{w}^{\left( q \right)}} \right)={{u}^{-1/2}}{{e}^{\sqrt{-1}{{w}^{\left( q \right)}}t}}={{u}^{-1/2}}{{e}^{\sqrt{-1}\frac{{{E}^{\left( q \right)}}}{\hbar }t}}
,~\alpha=1$$ where ${{E}^{\left( q \right)}}=\hbar{{w}^{\left( q \right)}} $ is the geometric energy eigenvalue.  Compare with the geometric eigenfunction, it obtains $\lambda=\sqrt{-1}{{w}^{\left( q \right)}}=\sqrt{-1}{{E}^{\left( q \right)}}/\hbar$, namely, $L\left( u,t,\sqrt{-1}w^{(q)}  \right)= F \left( u,t,{{w}^{\left( q \right)}} \right)$, therefore, the proof is completed.

\end{proof}
\end{corollary}
Note that the period $T=t$ is equal to the integral of time at the interval $ \left[ 0,t \right]$, and the continuous variable $u$ satisfies $u>0$.  Based on the geometric frequency eigenvalue ${{w}^{\left( q \right)}}$, it produces the geometric energy eigenvalue ${{E}^{\left( q \right)}}=\hbar{{w}^{\left( q \right)}} $ as well.

As a result of the theorem \ref{t2}, if we choose $f={u}^{-1/2}$ for \eqref{a2}, and then by using lemma \ref{l1}, it instantly gets
\begin{align}\label{b4}
  {{{\hat{w}}}^{\left( cl \right)}}\left( {{u}^{-1/2}}g \right)&={{u}^{-1/2}}{{{\hat{w}}}^{\left( cl \right)}}g-{{b}_{c}}{{u}^{-1/2}}g\frac{du}{dx} \\
 & ={{u}^{-1/2}}\left( {{{\hat{w}}}^{\left( cl \right)}}-{{b}_{c}}\frac{du}{dx} \right)g \notag\\
 & =2{{b}_{c}}{{u}^{1/2}}\frac{dg}{dx} \notag
\end{align}
Hence, we can deduce
\begin{equation}\label{a3}
  {{\hat{w}}^{\left( cl \right)}}\left( {{u}^{-1/2}}g \right)=2{{b}_{c}}\frac{{{u}^{1/2}}}{v}\frac{dg}{dt}=-\sqrt{-1}\alpha {{u}^{-1/2}}\frac{dg}{dt}
\end{equation}
where we have used $dx=vdt$ and $\hbar u=\alpha mv$, note that $\alpha \in \mathbb{R}$, in the most of time, we choose $\alpha=1$.
More precisely, it leads to a deformation of formula
\begin{align}
{{{\hat{w}}}^{\left( cl \right)}}\left( {{u}^{-1/2}}g \right)  &=2{{b}_{c}}{{u}^{1/2}}\frac{dg}{dx}=2{{b}_{c}}u\frac{d\ln g}{dx}\left( {{u}^{-1/2}}g \right)  \notag\\
 & =2{{b}_{c}}\frac{u}{v}\frac{d\ln g}{dt}\left( {{u}^{-1/2}}g \right)  \notag\\
 & =-\sqrt{-1}\alpha \lambda \left( {{u}^{-1/2}}g \right) \notag
\end{align}
where $\lambda =d\ln g/dt$ is used.
Therefore, \eqref{a2} can be rewritten as
\[{{\hat{w}}^{\left( cl \right)}}\left( fg \right)=f{{\hat{w}}^{\left( cl \right)}}g+g{{\hat{w}}^{\left( cl \right)}}f+\sqrt{-1}\alpha fg\frac{d\ln {{u}^{-1/2}}}{dt}\]and \eqref{a7} can be rewritten as
\[{{\hat{w}}^{\left( cl \right)}}{{c}_{1}}=\sqrt{-1}{{c}_{1}}\alpha \frac{d\ln {{u}^{-1/2}}}{dt}\]In particular, for \eqref{a7}, we have
${{\hat{w}}^{\left( cl \right)}}1=\sqrt{-1}\alpha \frac{d\ln {{u}^{-1/2}}}{dt}$.
In order to see \eqref{a3} easily in form, it can be rewritten as
\[\sqrt{-1}{{\hat{w}}^{\left( cl \right)}}\left( {{u}^{-1/2}}g \right)=\alpha {{u}^{-1/2}}\frac{dg}{dt}=\alpha \lambda\left( {{u}^{-1/2}}g \right)\]
where $\lambda =\frac{d\ln g}{dt}$ are the eigenvalues.
It's equivalent to the eigenvalues equation  $\frac{dg}{dt}=\lambda g$ which is the first order homogeneous linear differential equation such that
\[\sqrt{-1}{{\hat{w}}^{\left( cl \right)}}\left( {{u}^{-1/2}}g \right)=\lambda \left( {{u}^{-1/2}}g \right)\]
where $\alpha=1$ is chosen, $\lambda$ are the eigenvalues, then
$g\left( t \right)=g\left( 0 \right){{e}^{\lambda t}}$,
where $T=t-0=t$ is the period corresponding to the geometric frequency $\lambda$.

Based on \eqref{a2}, if we let $f=g={{u}^{-1/2}}$ be given, then
\[{{\hat{w}}^{\left( cl \right)}}{{u}^{-1}}=-\sqrt{-1}\alpha {{u}^{-1}}\frac{d\ln {{u}^{1/2}}}{dt}=-\frac{\sqrt{-1}\alpha }{2}\frac{d}{dt}{{u}^{-1}}\]can be deduced. Most importantly, ${{\hat{w}}^{\left( cl \right)}}$ \eqref{e1} can be expressed in another form
\[{{\hat{w}}^{\left( cl \right)}}={{b}_{c}}\left( 2u\frac{d}{dx}+\frac{du}{dx} \right)=-\sqrt{-1}\alpha \left( \frac{d}{dt}+\frac{d\ln {{u}^{1/2}}}{dt} \right)\]
where $\alpha =2{{b}_{c}}\sqrt{-1}\frac{u}{v}$, note that this expression says another essence of the G-dynamics ${{\hat{w}}^{\left( cl \right)}}$ with respect to the time variable.

For the geometric wave function \eqref{e4}, actually, we can verify it by using \eqref{b4}, if
we choose $g(x)=\exp \left( \sqrt{-1}a_{c}^{-1}\int{\frac{{{w}^{\left( q \right)}}}{u}dx} \right)$, then \[\frac{dg}{dx}=\sqrt{-1}a_{c}^{-1}g\frac{d}{dx}\int{\frac{{{w}^{\left( q \right)}}}{u}dx}=\sqrt{-1}a_{c}^{-1}g\frac{{{w}^{\left( q \right)}}}{u}=\sqrt{-1}{{\alpha }^{-1}}\frac{{{w}^{\left( q \right)}}}{v}g\]Subsequently,  \eqref{b4} precisely becomes
\[{{\hat{w}}^{\left( cl \right)}}\left( {{u}^{-1/2}}g \right)=2{{b}_{c}}{{u}^{1/2}}\frac{dg}{dx}={{w}^{\left( q \right)}}\left( {{u}^{-1/2}}g \right)\]
where $2\sqrt{-1}{{b}_{c}}a_{c}^{-1}=1$.

According to ${{\hat{w}}^{\left( cl \right)}}$ \eqref{e1}, the imaginary geomenergy follows
\[{{\hat{H}}^{\left( \operatorname{Im} \right)}}\left( {{{\hat{w}}}^{\left( cl \right)}} \right)=\sqrt{-1}\hbar {{\hat{w}}^{\left( cl \right)}}=\frac{{{\hbar }^{2}}}{2m}\left( 2u\frac{d}{dx}+\frac{du}{dx} \right)\]Then ${{\hat{H}}^{\left( cl \right)}}\left( {{{\hat{w}}}^{\left( cl \right)}} \right){{u}^{-1/2}}=0$.

As a consequence of the eigenvalues equation \eqref{a1} of the G-dynamics $\sqrt{-1}{{\hat{w}}^{\left( cl \right)}}$ given by theorem \ref{t1}, we have the following corollary about the imaginary geomenergy expressed as
\begin{corollary}
   The eigenvalues equation of the imaginary geomenergy  ${{\hat{H}}^{\left( \operatorname{Im} \right)}}\left( {{{\hat{w}}}^{\left( cl \right)}} \right)$ is given by
   \begin{equation}\label{c7}
     {{\hat{H}}^{\left( \operatorname{Im} \right)}}\left( {{{\hat{w}}}^{\left( cl \right)}} \right)L\left( u,t,\lambda  \right)={{E}^{\left( \operatorname{Im} \right)}}L\left( u,t,\lambda  \right)
   \end{equation}
where energy spectrum ${{E}^{\left( \operatorname{Im} \right)}}=\hbar \lambda $,  eigenvalues $\lambda$ in terms of geometric eigenfunction $L\left( u,t,\frac{{{E}^{\left( \operatorname{Im} \right)}}}{\hbar } \right)={{u}^{-1/2}}{{e}^{{{\frac{{{E}^{\left( \operatorname{Im} \right)}}}{\hbar } }}t}}$ are given.
\begin{proof}
By using the theorem \ref{t1}, we can derive eigenvalues equation of the imaginary geomenergy as follows
\begin{align}
  {{\hat{H}}^{\left( \operatorname{Im} \right)}}\left( {{{\hat{w}}}^{\left( cl \right)}} \right)L\left( u,t,\lambda  \right)& =\sqrt{-1}\hbar {{{\hat{w}}}^{\left( cl \right)}}L\left( u,t,\lambda  \right) \notag\\
 & =\lambda \hbar L\left( u,t,\lambda  \right) \notag\\
 & ={{E}^{\left( \operatorname{Im} \right)}}L\left( u,t,\lambda  \right) \notag
\end{align}
where ${{E}^{\left( \operatorname{Im} \right)}}=\hbar \lambda$, hence, we complete the proof.
\end{proof}
\end{corollary}

Consider the square form of the G-dynamics ${{\hat{w}}^{\left( cl \right)}}$ \eqref{e1}, by simply calculation, it yields
\begin{equation}\label{c5}
  {{\hat{w}}^{\left( cl \right)2}}=b_{c}^{2}\left( 4{{u}^{2}}\frac{{{d}^{2}}}{d{{x}^{2}}}+{{K}^{2}} \right)+\hat{\alpha }=4{{u}^{2}}b_{c}^{2}\left( \frac{{{d}^{2}}}{d{{x}^{2}}}+{{k}^{2}} \right)+\hat{\alpha }
\end{equation}
where $K\left( x \right)=\frac{du}{dx}$ has been used, $k\left( x \right)=\frac{K}{2u}=\frac{d\ln {{u}^{1/2}}}{dx}$, and
\begin{equation}\label{c4}
  \hat{\alpha }=2ub_{c}^{2}\left( 4K\frac{d}{dx}+\frac{dK}{dx} \right)
\end{equation}

\begin{theorem}
 For $\hat{\alpha }h(x)=0$ holds with $K>0$ if and only if
  $h(x)={{C}_{1}}{{K}^{-1/4}}$, where ${C}_{1}$ is constant.
\begin{proof}
  Based on  $\hat{\alpha }h(x)=0$ holds with $K>0$, it's equivalent to first order differential equation
  $2K{{h}_{x}}+{{K}_{x}}h=0$ for a given real-valued function $h$, then its solution is given by
  $h(x)={{C}_{1}}{{K}^{-1/4}}$, where ${C}_{1}$ is constant.
\end{proof}
\end{theorem}
As we notice, firstly, the expression of $\hat{\alpha }$ is analogous to the G-dynamics ${{\hat{w}}^{\left( cl \right)}}$ \eqref{e1}, the next analogous place is the solution as we can see.

According to the eigenvalues equation \eqref{a1}, it induces the equation
\[{{\hat{w}}^{\left( cl \right)2}}L\left( u,t,\lambda  \right)=-\sqrt{-1}\lambda {{\hat{w}}^{\left( cl \right)}}L\left( u,t,\lambda  \right)=-{{\lambda }^{2}}L\left( u,t,\lambda  \right)\]
As given, $L\left( u,t,\lambda  \right)$ is the eigenfunction of the operator ${{\hat{w}}^{\left( cl \right)2}}$ that is associated with eigenvalue $-{{\lambda }^{2}}$.

\section{Further case on the eigenvalues equation}
Let's consider geometrinetic energy operator ${{\hat{T}}^{\left( ri \right)}}$ similar to the square form of the G-dynamics \eqref{c5}  in one-dimensional case,  the geometrinetic energy operator ${{\hat{T}}^{\left( ri \right)}}$ as a geometric generalization of the classical kinetic energy operator is written as
\begin{equation}\label{c6}
  {{\hat{T}}^{\left( ri \right)}}=-\frac{{{\hbar }^{2}}}{2m}\left( {{d}^{2}}/{dx}^{2}+{{u}^{2}} \right)-\sqrt{-1}\hbar {{\hat{w}}^{\left( cl \right)}}
\end{equation}
where ${{\hat{E}}^{\left( w \right)}}=-\frac{{{\hbar }^{2}}}{2m}\frac{{{d}^{2}}}{d{{x}^{2}}}-\sqrt{-1}\hbar {{\hat{w}}^{\left( cl \right)}}$ is in one-dimensional motor operator, then geometrinetic energy operator is rewritten as
${{\hat{T}}^{\left( ri \right)}}={{\hat{E}}^{\left( w \right)}}-{{E}^{\left( s \right)}}/2$, where ${{E}^{\left( s \right)}}=\frac{{{\hbar }^{2}}}{m}{{u^{2}}}$ can be regarded as a potential energy.
If we consider eigenvalues equation of another one-dimensional operator concretely given by
\begin{align}
 {{{\hat{w}}}^{\left( ri \right)}}\Psi & =-\sqrt{-1}{{\left[ s,{{\hat{T}}^{\left( ri \right)}}+V(x) \right]}_{QPB}}\Psi /\hbar =-\sqrt{-1}{{\left[ s,{{\hat{E}}^{\left( w \right)}}-{{E}^{\left( s \right)}}/2+V\left( x \right) \right]}_{QPB}}\Psi /\hbar \notag\\
 & =-\frac{\sqrt{-1}}{\hbar }{{\left[ s,-\frac{{{\hbar }^{2}}}{2m}\left( {{d}^{2}}/d{{x}^{2}}+{{u}^{2}} \right)-\sqrt{-1}\hbar {{\hat{w}}^{\left( cl \right)}}+V\left( x \right) \right]}_{QPB}}\Psi  \notag\\
 & =-\frac{\sqrt{-1}}{\hbar }{{\left[ s,-\frac{{{\hbar }^{2}}}{2m}{{d}^{2}}/d{{x}^{2}}-\frac{{{\hbar }^{2}}}{2m}{{u}^{2}}-\sqrt{-1}\hbar {{\hat{w}}^{\left( cl \right)}}+V\left( x \right) \right]}_{QPB}}\Psi  \notag\\
 & =-\frac{\sqrt{-1}}{\hbar }{{\left[ s,-\frac{{{\hbar }^{2}}}{2m}{{d}^{2}}/d{{x}^{2}}-\sqrt{-1}\hbar {{\hat{w}}^{\left( cl \right)}} \right]}_{QPB}}\Psi \notag \\
 & =\frac{\sqrt{-1}\hbar }{2m}{{\left[ s,{{d}^{2}}/d{{x}^{2}} \right]}_{QPB}}\Psi -{{\left[ s,{{\hat{w}}^{\left( cl \right)}} \right]}_{QPB}}\Psi  \notag\\
 & ={{b}_{c}}\left( 2u{{\Psi }_{x}}+\Psi {{u}_{x}} \right)+2{{b}_{c}}{{u}^{2}}\Psi  \notag\\
 & ={{b}_{c}}\left( 2u{{\Psi }_{x}}+{{u}_{x}}\Psi +2{{u}^{2}}\Psi  \right) \notag\\
 &={{b}_{c}}\left( 2ud/dx+{{u}_{x}}+2{{u}^{2}} \right)\Psi \notag\\
 &={{\hat{w}}^{\left( cl \right)}}\Psi +2{{b}_{c}}{{u}^{2}}\Psi \notag\\
 &={{\hat{w}}^{\left( cl \right)}}\Psi +{{w}^{\left( s \right)}}\Psi \notag
\end{align}Then it yields ${{\hat{w}}^{\left( ri \right)}}={{\hat{w}}^{\left( cl \right)}}+{{w}^{\left( s \right)}}$,
where we have used \eqref{b6}, and the facts that ${{\left[ s,V\left( x \right)-{{E}^{\left( s \right)}}/2 \right]}_{QPB}}=0$ and antisymmetry 
\begin{equation}\label{e6}
  {{w}^{\left( s \right)}}=-{{\left[ s,{{\hat{w}}^{\left( cl \right)}} \right]}_{QPB}}={{\left[ {{\hat{w}}^{\left( cl \right)}},s \right]}_{QPB}}=2{{b}_{c}}{{u}^{2}}
\end{equation}
hold for the deduction above. Hence, it gets
$ {{\left[ s,{{{\hat{p}}}^{\left( cl \right)}} \right]}_{QPB}}=\sqrt{-1}\hbar u$, and ${{\left[ s,{{{\hat{H}}}^{\left( ri \right)}} \right]}_{QPB}}=\sqrt{-1}\hbar {{{\hat{w}}}^{\left( ri \right)}}$ together.
As a consequence of above derivation, we totally have non-Hermitian Hamiltonian operator given by
\begin{align}\label{e2}
{{\hat{H}}^{\left( ri \right)}}={{{\hat{T}}}^{\left( ri \right)}}+V\left( x \right)  &={{{\hat{E}}}^{\left( w \right)}}-{{E}^{\left( s \right)}}/2+V\left( x \right) ={{{\hat{H}}}^{\left( cl \right)}}-{{E}^{\left( s \right)}}/2-\sqrt{-1}\hbar {{\hat{w}}^{\left( cl \right)}} 
\end{align}
and using the formulation of the G-dynamics, it leads us to the certain result
\begin{align}
 {{\hat{w}}^{\left( ri \right)}} &=-\sqrt{-1}{{\left[ s,{{\hat{H}}^{\left( ri \right)}} \right]}_{QPB}}/\hbar =-\sqrt{-1}{{\left[ s,{{{\hat{E}}}^{\left( w \right)}}-{{E}^{\left( s \right)}}/2+V\left( x \right) \right]}_{QPB}}/\hbar   \notag\\
 & =-\sqrt{-1}{{\left[ s,{{{\hat{H}}}^{\left( cl \right)}}-{{E}^{\left( s \right)}}/2-\sqrt{-1}\hbar {{\hat{w}}^{\left( cl \right)}} \right]}_{QPB}}/\hbar  \notag\\
 & =-\sqrt{-1}{{\left[ s,{{{\hat{H}}}^{\left( cl \right)}}-\sqrt{-1}\hbar {{\hat{w}}^{\left( cl \right)}} \right]}_{QPB}}/\hbar   \notag\\
 & =-\sqrt{-1}{{\left[ s,{{{\hat{E}}}^{\left( w \right)}} \right]}_{QPB}}/\hbar  \notag
\end{align}
which is precisely shown by
\begin{equation}\label{b9}
  {{\hat{w}}^{\left( ri \right)}}={{b}_{c}}\left( 2ud/dx+{{u}_{x}}+2{{u}^{2}} \right)
\end{equation}
Actually, we have inserted the G-dynamics ${{\hat{w}}^{\left( cl \right)}} $ and curvature operator $\hat{Q}$ and ${{\hat{w}}^{\left( ri \right)}}$ into the \eqref{c6}, see what we have accomplished
\begin{align}
 {{{\hat{T}}}^{\left( ri \right)}}/\hbar &=-\sqrt{-1}{{b}_{c}}\left( {{d}^{2}}/d{{x}^{2}}+{{u}^{2}} \right)-\sqrt{-1}{{\hat{w}}^{\left( cl \right)}} =-\sqrt{-1}{{b}_{c}}\left( {{d}^{2}}/d{{x}^{2}}+{{u}^{2}}+\hat{Q} \right) \notag\\
 & =-\sqrt{-1}{{b}_{c}}\left( {{d}^{2}}/d{{x}^{2}}+{{u}^{2}}+2ud/dx+{{u}_{x}} \right) \notag\\
 & =-\sqrt{-1}{{b}_{c}}\left( {{d}^{2}}/d{{x}^{2}}-{{u}^{2}}+2{{u}^{2}}+2ud/dx+{{u}_{x}} \right) \notag\\
 & =-\sqrt{-1}{{b}_{c}}\left( {{d}^{2}}/d{{x}^{2}}-{{u}^{2}} \right)-\sqrt{-1}{{b}_{c}}\left( 2{{u}^{2}}+2ud/dx+{{u}_{x}} \right) \notag\\
 & =-\sqrt{-1}\left( {{b}_{c}}\left( {{d}^{2}}/d{{x}^{2}}-{{u}^{2}} \right)+{{\hat{w}}^{\left( ri \right)}} \right) \notag
\end{align}It has given the relation between the ${{\hat{w}}^{\left( ri \right)}}$ and geometrinetic energy operator ${{\hat{T}}^{\left( ri \right)}}$.
If we denote \[{{\hat{Q}}^{\left( c \right)}}={{d}^{2}}/d{{x}^{2}}+{{u}^{2}}+2ud/dx+{{u}_{x}}={{d}^{2}}/d{{x}^{2}}+{{u}^{2}}
+\hat{Q}\]
Then ${{\hat{T}}^{\left( ri \right)}}/\hbar =-\sqrt{-1}{{b}_{c}}{{\hat{Q}}^{\left( c \right)}}$. Notice that the characters of the ${{\hat{Q}}^{\left( c \right)}}$ is the free combination of the basic elements $d/dx,~u$, more specifically, namely, $$d/dx,~u\to {{d}^{2}}/d{{x}^{2}},~{{u}^{2}},~ud/dx,~{{u}_{x}}$$ The \eqref{b9} inspires us a relation between one-dimensional G-dynamics  ${{\hat{w}}^{\left( cl \right)}}$ and ${{\hat{w}}^{\left( ri\right)}}$,
\[{{\left[ s,{{{\hat{w}}}^{\left( ri \right)}} \right]}_{QPB}}={{\left[ s,{{{\hat{w}}}^{\left( cl \right)}}+{{w}^{\left( s \right)}} \right]}_{QPB}}={{\left[ s,{{{\hat{w}}}^{\left( cl \right)}} \right]}_{QPB}}=-2{{b}_{c}}{{u}^{2}}=-{{w}^{\left( s \right)}}\]
Making use of the identity \eqref{b7}, the same operation for ${{{\hat{Q}}}^{\left( c \right)}}$ shows us the result
\begin{align}\label{b8}
 {{{\hat{Q}}}^{\left( c \right)}}{{u}^{-1/2}} & ={{d}^{2}}{{u}^{-1/2}}/d{{x}^{2}}+{{u}^{2}}{{u}^{-1/2}}+\hat{Q}{{u}^{-1/2}} \notag\\
 & =-\frac{1}{2}{{u}^{-3/2}}\left( {{u}_{xx}}-\frac{3}{2}u_{x}^{2}{{u}^{-1}} \right)+{{u}^{3/2}} \\
 & =-\frac{1}{2}{{u}^{-3/2}}\left( {{u}_{xx}}-\frac{3}{2}u_{x}^{2}{{u}^{-1}}-2{{u}^{3}} \right) \notag
\end{align}where ${{d}^{2}}{{u}^{-1/2}}/d{{x}^{2}}=-{{u}^{-3/2}}\left( {{u}_{xx}}-\frac{3}{2}u_{x}^{2}{{u}^{-1}} \right)/2$ is given. Let's see how the ${{\hat{w}}^{\left( ri \right)}}{{u}^{-1/2}} $ in contrast to the identity ${{{\hat{w}}}^{\left( cl \right)}}{{u}^{-1/2}}=0$ becomes
\begin{align}
 {{\hat{w}}^{\left( ri \right)}}{{u}^{-1/2}} &={{b}_{c}}\left( 2{{u}^{2}}+2ud/dx+{{u}_{x}} \right){{u}^{-1/2}} \notag\\
 & ={{{\hat{w}}}^{\left( cl \right)}}{{u}^{-1/2}}+{{w}^{\left( s \right)}}{{u}^{-1/2}} \notag\\
 & ={{w}^{\left( s \right)}}{{u}^{-1/2}} \notag\\
 & =2{{b}_{c}}{{u}^{3/2}} \notag
\end{align}
As we did previously for the G-dynamics ${{\hat{w}}^{\left( cl \right)}} $, we also have ${{\hat{w}}^{\left( ri \right)}}1={{b}_{c}}\left( 2{{u}^{2}}+{{u}_{x}} \right)$, hence, we get ${{\hat{w}}^{\left( ri \right)}}{{c}_{1}}={{c}_{1}}{{\hat{w}}^{\left( ri \right)}}1$ immediately. And there are some special cases giving to us as follows,
\begin{align}
  & {{\hat{w}}^{\left( ri \right)}}x={{b}_{c}}\left( 2x{{u}^{2}}+2u+x{{u}_{x}} \right) \notag\\
 & {{\hat{w}}^{\left( ri \right)}}s={{b}_{c}}\left( \left( s+1 \right)2{{u}^{2}}+s{{u}_{x}} \right) \notag\\
 & {{\hat{w}}^{\left( ri \right)}}u={{b}_{c}}u\left( 2{{u}^{2}}+3{{u}_{x}} \right) \notag
\end{align}
Note that ${{\hat{w}}^{\left( ri \right)}}$ is also a linear operator, in particular, it satisfies
\begin{align}
{{\hat{w}}^{\left( ri \right)}}\left( fg \right)  &={{b}_{c}}\left( 2{{u}^{2}}+2ud/dx+{{u}_{x}} \right)\left( fg \right) \notag\\
 & ={{b}_{c}}\left( 2{{u}^{2}}fg+2ud\left( fg \right)/dx+fg{{u}_{x}} \right) \notag\\
 & =f{{\hat{w}}^{\left( ri \right)}}g+g{{\hat{w}}^{\left( ri \right)}}f-fg{{\hat{w}}^{\left( ri \right)}}1 \notag
\end{align}for two given functions $f,g$.
Analogically, ${{\hat{w}}^{\left( ri \right)}}$ does not satisfy the Leibniz identity, hence,
\begin{corollary}
  ${{\hat{w}}^{\left( ri \right)}}$ satisfies the Leibniz identity if and only if
  ${{\hat{w}}^{\left( ri \right)}}1 =0$.
\end{corollary}
By solving the condition  ${{\hat{w}}^{\left( ri\right)}}1=0$, it's equivalent to the condition ${{u}_{x}}+2{{u}^{2}}=0$, the solution of the ODE is equal to $u\left( x \right)=\frac{1}{2\left( x-{{x}_{0}} \right)}$, where ${{x}_{0}}$ is an initial value of the coordinate $x$, for convenient, we can choose the origin, that is  ${{x}_{0}}=0$ such that $u\left( x \right)=\frac{1}{2x},~x\neq 0$. Correspondingly, it derives $s=\ln {{x}^{1/2}}+{{C}_{10}},~x>0$, where ${{C}_{10}}$ is a constant.

In the beginning, we try to figure out whether there is the zero solution or not for operator ${{\hat{w}}^{\left( ri \right)}}$, that is if there exists a function $\Phi$ such that ${{\hat{w}}^{\left( ri \right)}}\Phi =0$ holds, namely, the ODE is given by
\[2u{{\Phi }_{x}}+\Phi {{u}_{x}}+2{{u}^{2}}\Phi =0\]By a direct evaluation, its solution can be obtained
$\Phi(x) ={{C}_{8}}{{u}^{-1/2}}{{e}^{-s}}$ accordingly. For a better approximation to $\Phi(x)$, we can fit a quadratic polynomial instead of a linear function:
\begin{align}
 \Phi(x) &={{C}_{8}}{{u}^{-1/2}}{{e}^{-s}}={{C}_{8}}{{u}^{-1/2}}-{{C}_{8}}s{{u}^{-1/2}}+{{C}_{8}}{{u}^{-1/2}}o\left( {{s}^{2}} \right) \notag
\end{align}
this polynomial has the same first and second derivatives, as is evident upon differentiation, where $o\left( {{s}^{2}} \right)$  is the approximation error.
Copying the operation of one-dimensional G-dynamics \eqref{e1} ${{\hat{w}}^{\left( cl \right)}}={{b}_{c}}\left( 2ud/dx+K \right)$, we seek the eigenfunction of the operator ${{\hat{w}}^{\left( ri \right)}}$ that is natural idea, it's described by the eigenvalues equation\[2u{{\phi }_{x}}+\phi {{u}_{x}}+2{{u}^{2}}\phi ={{w}^{\left( ri \right)}}\phi /{{b}_{c}} \]In effect, ${{\hat{w}}^{\left( ri \right)}}\phi ={{\hat{w}}^{\left( cl \right)}}\phi +{{w}^{\left( s \right)}}\phi ={{w}^{\left( cl \right)}}\phi +{{w}^{\left( s \right)}}\phi $, we denote ${{\hat{w}}^{\left( cl \right)}}\phi ={{w}^{\left( cl \right)}}\phi$ with respect to the eigenfunction $\phi$ and ${{w}^{\left( ri \right)}}={{w}^{\left( cl \right)}}+{{w}^{\left( s \right)}}$.  As a result, we obtain the eigenfunction
$\phi ={{C}_{9}}{{u}^{-1/2}}{{e}^{-s}}{{e}^{J\left( x \right)}}$, where
$J\left( x \right)=\int{\frac{{{w}^{\left( ri \right)}}}{2{{b}_{c}}u}dx}$ has been used, more precisely,
\begin{align}
 J\left( x \right) &=\int{\frac{{{w}^{\left( ri \right)}}}{2{{b}_{c}}u}dx}=\int{\frac{{{w}^{\left( cl \right)}}+{{w}^{\left( s \right)}}}{2{{b}_{c}}u}dx} =\int{\frac{{{w}^{\left( cl \right)}}}{2{{b}_{c}}u}dx}+\int{udx} \notag\\
 &=\frac{v}{2{{b}_{c}}u}\int{{{w}^{\left( cl \right)}}dt}+s \notag\\
 & =\sqrt{-1}{{\alpha }^{-1}}{{w}^{\left( cl \right)}}t+s \notag
\end{align}where $dx=vdt$ is used and $v$ is the one-dimensional velocity as well, $t=T=t-0$ is the period.
Then it leads to the eigenfunction which can be rewritten as
\begin{align}
  \phi& ={{C}_{9}}{{u}^{-1/2}}{{e}^{-s}}{{e}^{J\left( x \right)}}={{C}_{9}}{{u}^{-1/2}}{{e}^{-s}}{{e}^{\sqrt{-1}{{\alpha }^{-1}}{{w}^{\left( cl \right)}}t+s}} \notag\\
 & ={{C}_{9}}{{u}^{-1/2}}{{e}^{\sqrt{-1}{{\alpha }^{-1}}{{w}^{\left( cl \right)}}t}}\notag
\end{align}
As it shows, the solution is similar to the \eqref{e4}.  In fact, according to the eigenfunction \eqref{e4}, it results in an outcome
\[{{\hat{w}}^{\left( ri \right)}}F\left( u,{{w}^{\left( q \right)}} \right)={{\hat{w}}^{\left( cl \right)}}F\left( u,{{w}^{\left( q \right)}} \right)+{{w}^{\left( s \right)}}F\left( u,{{w}^{\left( q \right)}} \right)={{w}^{\left( q \right)}}F\left( u,{{w}^{\left( q \right)}} \right) +{{w}^{\left( s \right)}}F\left( u,{{w}^{\left( q \right)}} \right)\]This point reveals that ${{w}^{\left( q \right)}}={{w}^{\left( cl \right)}}$ with the same eigenfunction $\phi =\psi={{C}_{0}}F \left( u,{{w}^{\left( q \right)}} \right)$.
Actually, we can build a ODE model based on the \eqref{c6}, it's given by
${{\hat{T}}^{\left( ri \right)}}={{c}_{1}}\left( {{d}^{2}}/d{{x}^{2}}+{{u}^{2}} \right)-{{c}_{2}}{{\hat{w}}^{\left( cl \right)}}$,
where ${c}_{1},{c}_{2}$ are constants, then we have
\begin{align}
 {{{\hat{T}}}^{\left( ri \right)}}{{u}^{-1/2}} &={{c}_{1}}\left( {{d}^{2}}{{u}^{-1/2}}/d{{x}^{2}}+{{u}^{3/2}} \right)-{{c}_{2}}{{\hat{w}}^{\left( cl \right)}}{{u}^{-1/2}}\notag \\
 & =-\frac{{{c}_{1}}}{2}{{u}^{-3/2}}\left( {{u}_{xx}}-\frac{3}{2}u_{x}^{2}{{u}^{-1}}-2{{u}^{3}} \right) \notag
\end{align}where we have used lemma \ref{l1} and \eqref{b8} for above derivation.
For the Ri-Hamiltonian operator
${{\hat{H}}^{\left( ri \right)}}={{\hat{H}}^{\left( cl \right)}}-{{E}^{\left( s \right)}}/2-\sqrt{-1}\hbar {{\hat{w}}^{\left( cl \right)}}$.
Plugging the energy eigenvalues equation \eqref{c7} and \eqref{a1} into above Ri-Hamiltonian operator, see how the eigenvalues are, thusly, it naturally induces the result
\begin{align}
 {{{\hat{H}}}^{\left( ri \right)}}L\left( u,t,\lambda  \right) & ={{{\hat{H}}}^{\left( cl \right)}}L\left( u,t,\lambda  \right)-\frac{{{\hbar }^{2}}}{2m}{{u}^{2}}L\left( u,t,\lambda  \right)-\hbar \lambda L\left( u,t,\lambda  \right)  \notag\\
 & =-\frac{{{\hbar }^{2}}}{2m}\frac{{{d}^{2}}L\left( u,t,\lambda  \right)}{d{{x}^{2}}}-\left( \frac{{{\hbar }^{2}}}{2m}{{u}^{2}}+\lambda \hbar -V\left( x \right) \right)L\left( u,t,\lambda  \right)  \notag
\end{align}
As it displays, we mainly focus on the first part $-\frac{{{\hbar }^{2}}}{2m}\frac{{{d}^{2}}L\left( u,t,\lambda  \right)}{d{{x}^{2}}}$, therefore,
To start with the first derivative with respect to the coordinate is
$\frac{dL}{dx}={{L}_{x}}=\kappa L$,
where $\kappa ={{\kappa }_{1}}-\frac{1}{2}{{u}^{-1}}{{u}_{x}}={{\kappa }_{1}}-{{\left( \ln {{u}^{1/2}} \right)}_{x}}$ and ${{\kappa }_{1}}=\frac{\lambda }{v}$, subsequently, we obtain
$\frac{{{d}^{2}}L}{d{{x}^{2}}}={{L}_{xx}}=({{\kappa }^{2}}+{{\kappa }_{x}})L$.
Totally, it yields
\[{{\hat{H}}^{\left( ri \right)}}L=-\left( \frac{{{\hbar }^{2}}}{2m}\left( {{\kappa }^{2}}+{{\kappa }_{x}} \right)+\frac{{{\hbar }^{2}}}{2m}{{u}^{2}}+\lambda \hbar -V\left( x \right) \right)L={{E}^{\left( ri \right)}}L\]
More precisely, if we take $\lambda =\sqrt{-1}{{w}^{\left( q \right)}}$ into account above, then $\kappa ={{\kappa }_{1}}-{{\left( \ln {{u}^{1/2}} \right)}_{x}},{{\kappa }_{1}}=\sqrt{-1}\frac{{{w}^{\left( q \right)}}}{v}$, and
\[{{E}^{\left( ri \right)}}=-\left( \frac{{{\hbar }^{2}}}{2m}\left( {{\kappa }^{2}}+{{\kappa }_{x}} \right)+\frac{{{\hbar }^{2}}}{2m}{{u}^{2}}+\sqrt{-1}\hbar {{w}^{\left( q \right)}}-V\left( x \right) \right)\]

\section{Properties of geometric eigenfunction}
In this section, we mainly give a proof of one of properties of geometric eigenfunction.
Furthermore, for geometric eigenfunction
\begin{equation}\label{d2}
  L\left( u,t,{{\lambda }} \right)={{u}^{-1/2}}{{e}^{{{\lambda }}t}}
\end{equation}with ${{\hat{w}}^{\left( cl \right)}}{{u}^{-1/2}}\equiv0$
based on the eigenvalue equation
$${{\hat{w}}^{\left( cl \right)}}L\left( u,t,{{\lambda }} \right)=-\sqrt{-1}{{\lambda }}L\left( u,t,{{\lambda }} \right)$$   where ${\lambda }$ is the frequency eigenvalues of the G-dynamics with respect to the eigenfunction  $L\left( u,t,{{\lambda }} \right)$.
The eigenfunction \eqref{d2} which can be rewritten as  $L\left( u,t,{{\lambda }} \right)=L\left( u,0,{{\lambda }} \right)L\left( 1,t,{{\lambda }} \right)$, where $L\left( u,0,{{\lambda }} \right)=L\left( u,t,0 \right)=L\left( u,0,0 \right)={{u}^{-1/2}}$ and  $L\left( 1,t,{{\lambda }} \right)={{e}^{{{\lambda }}t}}$,  it implies that geometric eigenfunction \eqref{d2} has a different mode under different condition. For instance, for any nonzero function $f_{3}={{f}_{1}}{{f}_{2}}$ with ${{f}_{1}},{{f}_{2}}>0$, by using the mode of the eigenfunction \eqref{d2}, it can be rewritten as
$f_{3}={{f}_{1}}^{3/2}L\left( {{f}_{1}},t,\frac{\ln {{f}_{2}}}{t} \right),~~t\ne 0$.
Another example is exponential function given by
${{r}^{-z_{1}}}=L\left( {{r}^{2x_{1}}},t,-\sqrt{-1}\frac{\ln r}{t}y_{1} \right)$, where $r,t>0$, and complex variable $z_{1}=x_{1}+\sqrt{-1}y_{1}$,

Hence, as a result of the geometric eigenfunction \eqref{d2}, the eigenvalues equation is rewritten as
\begin{align}
  & \sqrt{-1}{{{\hat{w}}}^{\left( cl \right)}}L\left( u,t,{{\lambda }} \right)={{\lambda }}L\left( u,0,{{\lambda }} \right)L\left( 1,t,{{\lambda }} \right) \notag\\
 & {{{\hat{w}}}^{\left( cl \right)}}L\left( u,0,{{\lambda }} \right)=0 \notag
\end{align}
Now, we can turn to analyze the properties of the eigenfunction $L\left( u,t,{{\lambda }} \right)$ as follows:
\[L\left( {{u}_{1}},{{t}_{1}},{{\lambda }_{1}} \right)L\left( {{u}_{2}},{{t}_{2}},{{\lambda }_{2}} \right)=L\left( {{u}_{1}}{{u}_{2}},{{t}_{1}}+{{t}_{2}},{{\lambda }_{2}} \right)L\left( 1,{{t}_{1}},{{\lambda }_{1}}-{{\lambda }_{2}} \right)\]
and $L\left( {{u}_{1}}{{u}_{2}},{{t }_{1}}+{{t}_{2}},\lambda\right)=L\left( {{u}_{1}},{{t }_{1}},{{\lambda }} \right)L\left( {{u}_{2}},{t}_{2},{{\lambda }} \right)$.
More generally, it leads to the following consequences which can be proven as follows:
\begin{theorem}
  The geometric eigenfunction has following properties given by
  \begin{equation}\label{cc1}
  B\left( \lambda  \right)=L\left( \prod\limits_{j}{{{u}_{j}}},\sum\limits_{j}{{{t}_{j}}},\lambda  \right)=\prod\limits_{j}{L\left( {{u}_{j}},{{t}_{j}},\lambda  \right)}
\end{equation}
where $L\left( {{u}_{j}},{{t}_{j}},\lambda  \right)=u_{j}^{-1/2}{{e}^{\lambda {{t}_{j}}}},~~j=1,2,\cdots \infty$ and ${t}_{j}$ is the primitive period of the orbit, and $Z\left( \lambda  \right)=\sum\limits_{j}{L\left( {{u}_{j}},{{t}_{j}},\lambda  \right)}$ follows.
\begin{proof}
According to the formula of the geometric eigenfunction \eqref{d2}, it directly derives
\begin{align}
  B\left( \lambda  \right)& =L\left( \prod\limits_{j}{{{u}_{j}}},\sum\limits_{j}{{{t}_{j}}},\lambda  \right)={{\left( \prod\limits_{j}{{{u}_{j}}} \right)}^{-1/2}}{{e}^{\lambda \sum\limits_{j}{{{t}_{j}}}}}=\left( \prod\limits_{j}{{{u}_{j}}^{-1/2}} \right)\prod\limits_{j}{{{e}^{\lambda {{t}_{j}}}}} \notag\\
 & =\prod\limits_{j}{{{u}_{j}}^{-1/2}{{e}^{\lambda {{t}_{j}}}}}=\prod\limits_{j}{L\left( {{u}_{j}},{{t}_{j}},\lambda  \right)},~~j=1,2,\cdots \infty \notag
\end{align}
where $L\left( {{u}_{j}},{{t}_{j}},\lambda  \right)=u_{j}^{-1/2}{{e}^{\lambda {{t}_{j}}}}$ and ${t}_{j}$ is the primitive period of the orbit. Meanwhile,  $Z\left( \lambda  \right)=\sum\limits_{j}{L\left( {{u}_{j}},{{t}_{j}},\lambda  \right)}$ just follows.
\end{proof}
\end{theorem}

\subsection{Coordinate derivative of the G-dynamics}
In the next, we give the commutator and generalized geometric commutator for the one dimensional G-dynamics \eqref{e1} ${{{\hat{w}}}^{\left( cl \right)}}$ and a differential function $q=q\left( x \right)$ respectively, namely,
${{\left[ {{{\hat{w}}}^{\left( cl \right)}},q\left( x \right) \right]}_{QPB}}=2{{b}_{c}}u{{q}_{x}}$, where $dq/dx={{q}_{x}}$ and
\[\left[ {{{\hat{w}}}^{\left( cl \right)}},q\left( x \right) \right]={{\left[ {{{\hat{w}}}^{\left( cl \right)}},q\left( x \right) \right]}_{QPB}}+G\left( s,{{{\hat{w}}}^{\left( cl \right)}},q\left( x \right) \right)=2{{b}_{c}}u\left( q_{x}+uq \right)\]where the quantum geometric
bracket for them is $G\left( s,{{{\hat{w}}}^{\left( cl \right)}},q\left( x \right) \right)=2{{b}_{c}}{{u}^{2}}q$.
Clearly, if the covariant equilibrium equation
$\left[ {{{\hat{w}}}^{\left( cl \right)}},q\left( x \right) \right]=0$ holds, then ${{q}_{x}}=-uq$, the solution $q={{C}_{2}}{{e}^{-s}}$ follows, ${{C}_{2}}$ is a constant. In particular, we choose $q=s$, then
${{\left[ {{{\hat{w}}}^{\left( cl \right)}},s\left( x \right) \right]}_{QPB}}=2{{b}_{c}}u^{2}$, and ${{\left[ {{{\hat{w}}}^{\left( cl \right)}},c_{1} \right]}_{QPB}}=0$, where $c_{1}$ is a constant.

 To construct an operator called acceleration operator in a form is
\begin{align}
 \hat{a}_{2} &={{{\hat{w}}}^{\left( cl \right)}}{{\hat{v}}^{\left( cl \right)}}=u{{\hat{v}}^{\left( cl \right)2}}+\frac{1}{2}({{\hat{v}}^{\left( cl \right)}}u){{\hat{v}}^{\left( cl \right)}}=2b_{c}^{2}\left( 2u{{d}^{2}}/d{{x}^{2}}+{{u}_{x}}d/dx \right) \notag
\end{align}
and G-dynamics \eqref{e1} is ${{\hat{w}}^{\left( cl \right)}}={{b}_{c}}\left( 2u\frac{d}{dx}+{{u}_{x}} \right)={{b}_{c}}\hat{Q}$,  where $\hat{Q}=2u\frac{d}{dx}+{{u}_{x}}$.
Hence, we consider its eigenvalues equation
$\hat{a}_{2}\phi ={{\lambda }_{a}}\phi $.  It can be rewritten as a standard second order differential equation \[{{\phi }_{xx}}+\frac{{{u}_{x}}}{2u}{{\phi }_{x}}-\frac{{{\lambda }_{a}}}{4ub_{c}^{2}}\phi =0\]
Simply, we consider $\hat{a}_{2}\phi =0$, namely, the equation
$2u{{\phi }_{xx}}+{{u}_{x}}{{\phi }_{x}}=0$, actually, we can obtain ${{\phi }_{x}}={{C}_{3}}{{u}^{-1/2}}$, where ${C}_{3}$ is a constant, then it leads to a solution $\phi(x) ={{C}_{3}}\int{{{u}^{-1/2}}dx}$.

Meanwhile, in a similar construction, we get
\begin{align}
 {{{\hat{a}}}_{1}} & ={{{\hat{v}}}^{\left( cl \right)}}{{{\hat{w}}}^{\left( cl \right)}}={{{\hat{v}}}^{\left( cl \right)}}u{{{\hat{v}}}^{\left( cl \right)}}+\frac{1}{2}{{{\hat{v}}}^{\left( cl \right)2}}u \notag\\
 & =2b_{c}^{2}\left( 2u{{d}^{2}}/d{{x}^{2}}+3{{u}_{x}}d/dx+{{u}_{xx}} \right) \notag\\
 & =-\frac{{{\hbar }^{2}}}{2{{m}^{2}}}\left( 2u{{d}^{2}}/d{{x}^{2}}+3{{u}_{x}}d/dx+{{u}_{xx}} \right) \notag
\end{align}then it yields geometorce operator with respect to the G-dynamics ${{\hat{w}}^{\left( cl \right)}} $
\[{{\hat{F}}^{\left( \operatorname{Im} \right)}}\left( x \right)=\frac{{{\hbar }^{2}}}{2m}\left( 2u\frac{{{d}^{2}}}{d{{x}^{2}}}+3{{u}_{x}}\frac{d}{dx}+{{u}_{xx}} \right)=\sqrt{-1}\hbar \frac{d{{{\hat{w}}}^{\left( cl \right)}}}{dx}=-m{{\hat{a}}_{1}}\]and conversely, we get ${{\hat{a}}_{1}}=2{{b}_{c}}\frac{d{{{\hat{w}}}^{\left( cl \right)}}}{dx}=-{{\hat{F}}^{\left( \operatorname{Im} \right)}}\left( x \right)/m$,
hence, it derives $d{{\hat{w}}^{\left( cl \right)}}/dx={{\hat{a}}_{1}}/2{{b}_{c}}=-{{\hat{F}}^{\left( \operatorname{Im} \right)}}\left( x \right)/\left( 2{{b}_{c}}m \right)$.
We can prove that \[\frac{d{{{\hat{w}}}^{\left( cl \right)}}}{dx}\Phi ={{b}_{c}}\frac{d}{dx}\left( 2u{{\Phi }_{x}}+\Phi {{u}_{x}} \right)={{b}_{c}}\left( 2u\frac{{{d}^{2}}}{d{{x}^{2}}}+3{{u}_{x}}\frac{d}{dx}+{{u}_{xx}} \right)\Phi \]
As a result, it has \[{{\hat{F}}^{\left( \operatorname{Im} \right)}}\left( x \right)\Phi =\frac{{{\hbar }^{2}}}{2m}\left( 2u{{\Phi }_{xx}}+3{{u}_{x}}{{\Phi }_{x}}+\Phi {{u}_{xx}} \right)=\sqrt{-1}\hbar \frac{d{{{\hat{w}}}^{\left( cl \right)}}}{dx}\Phi =-m{{\hat{a}}_{1}}\Phi \]
If let ${{\hat{a}}_{1}}\Phi =0$ be given, then $d{{\hat{w}}^{\left( cl \right)}}/dx=0$ and ODE is obtained basically
\[2u{{\Phi }_{xx}}+3{{u}_{x}}{{\Phi }_{x}}+\Phi {{u}_{xx}}=0\]or the standard form of the ODE ${{\Phi }_{xx}}+\frac{3{{u}_{x}}}{2u}{{\Phi }_{x}}+\Phi \frac{{{u}_{xx}}}{2u}=0$. Similarly, the coordinate derivative of the ${{\hat{w}}}^{\left( ri \right)}$ is given by 
\begin{align}
 \frac{d{{{\hat{w}}}^{\left( ri \right)}}}{dx}\Phi & ={{b}_{c}}\frac{d}{dx}\left( 2u{{\Phi }_{x}}+\Phi {{u}_{x}}+2\Phi {{u}^{2}} \right) \notag\\
 & ={{b}_{c}}\left( 2u\frac{{{d}^{2}}}{d{{x}^{2}}}+3{{u}_{x}}\frac{d}{dx}+{{u}_{xx}} \right)\Phi +2{{b}_{c}}\left( {{\Phi }_{x}}{{u}^{2}}+2u{{u}_{x}}\Phi  \right) \notag\\
 & ={{b}_{c}}\left( 2u\left( \frac{{{d}^{2}}}{d{{x}^{2}}}+2{{u}_{x}} \right)+\left( 3{{u}_{x}}+2{{u}^{2}} \right)\frac{d}{dx}+{{u}_{xx}} \right)\Phi  \notag
\end{align}
As the formula shows above, it directly gives the result \[\frac{d{{{\hat{w}}}^{\left( ri \right)}}}{dx}=\frac{d{{{\hat{w}}}^{\left( cl \right)}}}{dx}+2{{b}_{c}}\left( {{u}^{2}}\frac{d}{dx}+2u{{u}_{x}} \right)\]It leads to another geometorce operator that is expressed as 
\begin{align}
 {{{\hat{F}}}^{\left( ri \right)}}\left( x \right) &=\sqrt{-1}\hbar \frac{d{{{\hat{w}}}^{\left( ri \right)}}}{dx}=\sqrt{-1}\hbar \frac{d{{{\hat{w}}}^{\left( cl \right)}}}{dx}+2\sqrt{-1}\hbar {{b}_{c}}\left( {{u}^{2}}\frac{d}{dx}+2u{{u}_{x}} \right) \notag\\
 & ={{{\hat{F}}}^{\left( \operatorname{Im} \right)}}\left( x \right)+{{{\hat{F}}}^{\left( s \right)}}\left( x \right) \notag
\end{align}
where \[{{\hat{F}}^{\left( s \right)}}\left( x \right)=2\sqrt{-1}\hbar {{b}_{c}}\left( {{u}^{2}}\frac{d}{dx}+2u{{u}_{x}} \right)=\frac{{{\hbar }^{2}}u}{m}\left( u\frac{d}{dx}+2{{u}_{x}} \right)\]
Note that there are three various kinds of the geometorce operators ${{\hat{F}}^{\left( ri \right)}}\left( x \right)$,${{\hat{F}}^{\left( \operatorname{Im} \right)}}\left( x \right)$, and ${{\hat{F}}^{\left( s \right)}}\left( x \right)$ satisfying the relation above, it mainly explains the derivative of the G-dynamics in terms of the coordinates. If   $\frac{d{{{\hat{w}}}^{\left( ri \right)}}}{dx}\Phi =0$, then \[2u{{\Phi }_{xx}}+\left( 3{{u}_{x}}+2{{u}^{2}} \right){{\Phi }_{x}}+\left( {{u}_{xx}}+4u{{u}_{x}} \right)\Phi =0\]
According to the non-Hermitian Hamiltonian operators ${{\hat{H}}}^{\left( ri \right)}$ \eqref{e2},  copying the same mode for the ${{\hat{H}}^{\left( ri \right)}}={{\hat{H}}^{\left( cl \right)}}-\frac{{{\hbar }^{2}}}{2m}{{u}^{2}}-\sqrt{-1}\hbar {{\hat{w}}^{\left( cl \right)}}$ \eqref{e2} leads to the a force operator that is similar to the geometorce operator
\begin{align}\label{e5}
 {{\hat{F}}^{\left( r \right)}} & =-\frac{d{{{\hat{H}}}^{\left( ri \right)}}}{dx}=-\frac{d}{dx}\left( {{{\hat{H}}}^{\left( cl \right)}}-\frac{{{\hbar }^{2}}}{2m}{{u}^{2}}-\sqrt{-1}\hbar {{{\hat{w}}}^{\left( cl \right)}} \right) \notag\\
 & =-\frac{d}{dx}{{{\hat{H}}}^{\left( cl \right)}}+\frac{d}{dx}{{E}^{\left( s \right)}}/2+\sqrt{-1}\hbar \frac{d}{dx}{{{\hat{w}}}^{\left( cl \right)}} \notag\\
 & =-\frac{d}{dx}{{{\hat{H}}}^{\left( cl \right)}}+\frac{{{\hbar }^{2}}}{m}u{{u}_{x}}+\frac{{{\hbar }^{2}}}{2m}{{u}^{2}}\frac{d}{dx}+\sqrt{-1}\hbar \frac{d}{dx}{{{\hat{w}}}^{\left( cl \right)}} \\
 & =\frac{{{\hbar }^{2}}}{2m}\frac{{{d}^{3}}}{d{{x}^{3}}}-{{V}_{x}}+\left( \frac{{{\hbar }^{2}}}{2m}{{u}^{2}}-V \right)\frac{d}{dx}+\frac{{{\hbar }^{2}}}{2m}\left( 2u\frac{{{d}^{2}}}{d{{x}^{2}}}+{{u}_{x}}\left( 2u+3\frac{d}{dx} \right)+{{u}_{xx}} \right) \notag\\
 & =\frac{{{\hbar }^{2}}}{m}{{u}_{x}}u-{{V}_{x}}+\left( \frac{{{\hbar }^{2}}}{2m}{{u}^{2}}-V \right)\frac{d}{dx}+\frac{{{\hbar }^{2}}}{2m}\left( \frac{{{d}^{3}}}{d{{x}^{3}}}+2u\frac{{{d}^{2}}}{d{{x}^{2}}}+3{{u}_{x}}\frac{d}{dx}+{{u}_{xx}} \right) \notag
\end{align}
If let ${{E}^{\left( s \right)}}/2=\frac{{{\hbar }^{2}}}{2m}{{u}^{2}}=V$ be given, then the non-Hermitian Hamiltonian operators ${{\hat{H}}}^{\left( ri \right)}$ \eqref{e2} becomes a simple form  ${{\hat{H}}^{\left( ri \right)}}={{\hat{H}}^{\left( cl \right)}}-\sqrt{-1}\hbar {{\hat{w}}^{\left( cl \right)}}$ and force operator \eqref{e5} is simply rewritten as
\[{{\hat{F}}^{\left( r \right)}}=-\frac{d{{{\hat{H}}}^{\left( ri \right)}}}{dx}=\frac{{{\hbar }^{2}}}{2m}\left( \frac{{{d}^{3}}}{d{{x}^{3}}}+2u\frac{{{d}^{2}}}{d{{x}^{2}}}+3{{u}_{x}}\frac{d}{dx}+{{u}_{xx}} \right)\]Therefore, when it comes to a function $\Phi$, it yields 
\[{{\hat{F}}^{\left( r \right)}}\Phi =\frac{{{\hbar }^{2}}}{2m}\left( {{\Phi }_{xxx}}+2u{{\Phi }_{xx}}+3{{u}_{x}}{{\Phi }_{x}}+\Phi {{u}_{xx}} \right)\]
If we take condition ${{\hat{F}}^{\left( r \right)}}\Phi=0 $ into account, then it deduces the third order ODE that is equal to 
\[{{\Phi }_{xxx}}+2u{{\Phi }_{xx}}+3{{u}_{x}}{{\Phi }_{x}}+\Phi {{u}_{xx}}=0\]
As a consequence, by a difference between ${{\hat{a}}_{1}}$ and ${{\hat{a}}_{2}}$, we obtain
\begin{align}\label{c3}
 \Delta \hat{a} &={{{\hat{a}}}_{1}}-{{{\hat{a}}}_{2}}={{\left[ {{{\hat{v}}}^{\left( cl \right)}},{{{\hat{w}}}^{\left( cl \right)}} \right]}_{QPB}}=2b_{c}^{2}\left( 2{{u}_{x}}\frac{d}{dx}+{{u}_{xx}} \right)\\
 & =2b_{c}^{2}\left( 2K\frac{d}{dx}+K_{x} \right) \notag
\end{align}
where $K_{x}=dK/dx={{u}_{xx}}={{d}^{2}}u/d{{x}^{2}}$ has been used. in general, $\Delta \hat{a}\neq 0$, since ${{\left[ {{{\hat{v}}}^{\left( cl \right)}},{{{\hat{w}}}^{\left( cl \right)}} \right]}_{QPB}}\neq 0$ holds generally, then that's why there are two kinds of acceleration operator.

\begin{theorem}
  $\Delta \hat{a}= 0$ holds if and only if ${{f}_{1}}={{C}_{4}}K^{-1/2}$ for a real-valued function $f_{1}$, where ${{C}_{4}}$ is a constant.
  \begin{proof}
    By solving the equation $2{{u}_{x}}{{f_{1}}_{x}}+f_{1}{{u}_{xx}}=0$, it easily verifies that ${{f}_{1}}={{C}_{4}}K^{-1/2}$  is the unique solution, where $K=u_{x}$ and ${{C}_{4}}$ is a constant.
\end{proof}
\end{theorem}
According to the generalized Heisenberg equation, we can obtain the time evolution
\begin{align}\label{d1}
 d{{\hat{v}}^{\left( cl \right)}}/dt &=\frac{1}{\sqrt{-1}\hbar }\left( {{\left[ {{\hat{v}}^{\left( cl \right)}},{{\hat{H}}^{\left( cl \right)}} \right]}_{QPB}}-{{\hat{H}}^{\left( cl \right)}}{{\left[ s,{{\hat{v}}^{\left( cl \right)}} \right]}_{QPB}} \right)=-\frac{1}{m}\left( {{V}_{x}}+{{\hat{H}}^{\left( cl \right)}}u \right)
\end{align}in terms of the velocity operator ${\hat{v}}^{\left( cl \right)}$, where ${{\left[ s,{{\hat{v}}^{\left( cl \right)}} \right]}_{QPB}}=-2b_{c}u$. More precisely, by direct computations, the equation \eqref{d1} can be specifically written as \[d{{\hat{v}}^{\left( cl \right)}}/dt=-\left( {{V}_{x}}+Vu \right)/m+\frac{{{\hbar }^{2}}}{2{{m}^{2}}}\left( u\frac{{{d}^{2}}}{d{{x}^{2}}}+2{{u}_{x}}\frac{d}{dx}+{{u}_{xx}} \right)\]
Then the covariant dynamics of a dynamical velocity variable ${{\hat{v}}^{\left( cl \right)}}$ follows
\begin{align}
 \frac{\mathcal{D}{{\hat{v}}^{\left( cl \right)}}}{dt} &=\frac{d{{\hat{v}}^{\left( cl \right)}}}{dt}+\hat{a}_{1}=-\frac{{{\hbar }^{2}}}{2{{m}^{2}}}\left( u\frac{{{d}^{2}}}{d{{x}^{2}}}+{{u}_{x}}\frac{d}{dx} \right)-\frac{1}{m}\left( {{V}_{x}}+uV\left( x \right) \right)  \notag
\end{align}
Notice that if we consider Hamiltonian operator ${{\hat{H}}^{\left( cl \right)}}$ without the potential energy $V$, namely, $V=0$, above covariant dynamics of a dynamical variable ${{\hat{v}}^{\left( cl \right)}}$ can be simplified as
$$\frac{\mathcal{D}{{\hat{v}}^{\left( cl \right)}}}{dt}=-\frac{{{\hbar }^{2}}}{2{{m}^{2}}}\left( u\frac{{{d}^{2}}}{d{{x}^{2}}}+{{u}_{x}}\frac{d}{dx} \right)$$
The quantum covariant equilibrium equation is given by
$\frac{\mathcal{D}{{\hat{v}}^{\left( cl \right)}}}{dt}=0$, hence, we can have an equation given by
$u{{f}_{xx}}+{{u}_{x}}{{f}_{x}}=0$ in terms of the function $f$, it gets
${{f}_{x}}={{C}_{5}}{{u}^{-1}}$ by solving the equation, furthermore, we obtain a solution $$f={{C}_{5}}\int{{{u}^{-1}}}dx$$ In fact, if the potential energy satisfies $V={{C}_{6}}{{e}^{-s}}$, where ${{C}_{5}},{{C}_{6}}$ are constants, then ${{V}_{x}}+uV\left( x \right)=0$ holds for the same effect at $V=0$ for covariant dynamics of a dynamical variable ${{\hat{v}}^{\left( cl \right)}}$, this can be verified, basically.

\subsection{Time evolution of the G-dynamics}
In this section, as a quantum dynamics, searching for the time evolution defined by  the generalized Heisenberg equation of the G-dynamics is needed as well.
Copying the time evolution of the equation \eqref{d1} for the G-dynamics ${{\hat{w}}^{\left( cl \right)}}$ leads us to the equation
\[d{{\hat{w}}^{\left( cl \right)}}/dt=\frac{1}{\sqrt{-1}\hbar }\left( {{\left[ {{{\hat{w}}}^{\left( cl \right)}},{{{\hat{H}}}^{\left( cl \right)}} \right]}_{QPB}}-{{{\hat{H}}}^{\left( cl \right)}}{{\left[ s,{{{\hat{w}}}^{\left( cl \right)}} \right]}_{QPB}} \right)\]More precisely, we give a complete deduction of the time evolution of the one-dimensional G-dynamics ${{\hat{w}}^{\left( cl \right)}}$ based on the formula of the generalized Heisenberg equation that is computed as
\begin{align}
 \frac{d}{dt}\left( {{{\hat{w}}}^{\left( cl \right)}}\Psi  \right) &=\frac{1}{\sqrt{-1}\hbar }\left( {{\left[ {{{\hat{w}}}^{\left( cl \right)}},{{{\hat{H}}}^{\left( cl \right)}} \right]}_{QPB}}\Psi -{{{\hat{H}}}^{\left( cl \right)}}{{\left[ s,{{{\hat{w}}}^{\left( cl \right)}} \right]}_{QPB}}\Psi  \right) \notag\\
 & =\frac{1}{\sqrt{-1}\hbar }\left( {{\left[ {{{\hat{w}}}^{\left( cl \right)}},{{{\hat{H}}}^{\left( cl \right)}} \right]}_{QPB}}\Psi +{{{\hat{H}}}^{\left( cl \right)}}{{\left[ {{{\hat{w}}}^{\left( cl \right)}},s \right]}_{QPB}}\Psi  \right) \notag\\
 & =\frac{2{{b}_{c}}}{\sqrt{-1}\hbar }\left( \frac{{{\hbar }^{2}}}{m}\left( \frac{m}{{{\hbar }^{2}}}u{{V}_{x}}+{{u}_{x}}\frac{{{d}^{2}}}{d{{x}^{2}}}+{{u}_{xx}}\frac{d}{dx}+\frac{1}{4}{{u}_{xxx}} \right)\Psi +{{{\hat{H}}}^{\left( cl \right)}}{{u}^{2}}\Psi  \right) \notag\\
 & =-\frac{{{\hbar }^{2}}}{{{m}^{2}}}\left( \frac{m}{{{\hbar }^{2}}}u{{V}_{x}}+{{u}_{x}}\frac{{{d}^{2}}}{d{{x}^{2}}}+{{u}_{xx}}\frac{d}{dx}+\frac{1}{4}{{u}_{xxx}} \right)\Psi  \notag\\
 & \begin{matrix}
   {} & {} & {} & {} & {}& {} & {} \\
\end{matrix}+\frac{{{\hbar }^{2}}}{{{m}^{2}}}\left( 2u{{u}_{x}}\frac{d}{dx}+{{u}_{x}}^{2}+u{{u}_{xx}}+{{u}^{2}}\frac{{{d}^{2}}}{2d{{x}^{2}}} \right)\Psi -\frac{{{u}^{2}}}{m}V\Psi  \notag\\
 & =-\frac{{{\hbar }^{2}}}{{{m}^{2}}}\left( \left( {{u}_{x}}-\frac{{{u}^{2}}}{2} \right)\frac{{{d}^{2}}}{d{{x}^{2}}}+\left( {{u}_{xx}}-2u{{u}_{x}} \right)\frac{d}{dx}+\frac{1}{4}{{u}_{xxx}}-{{u}_{x}}^{2}-u{{u}_{xx}} \right)\Psi  \notag\\
 & \begin{matrix}
   {} & {} & {} & {} & {}& {} & {} \\
\end{matrix}-\frac{1}{m}\left( u{{V}_{x}}+{{u}^{2}}V \right)\Psi  \notag
\end{align}
where the relation \eqref{e6} has put into the derivation above, and accordingly, by direct evaluation, it derives  \[{{\hat{H}}^{\left( cl \right)}}{{u}^{2}}\Psi =-\frac{{{\hbar }^{2}}}{2m}\left( 4u{{u}_{x}}\frac{d}{dx}+2{{u}_{x}}^{2}+2u{{u}_{xx}}+{{u}^{2}}\frac{{{d}^{2}}}{d{{x}^{2}}} \right)\Psi +{{u}^{2}}V\Psi \]and the commutator of the G-dynamics ${{{\hat{w}}}^{\left( cl \right)}}$ and the classical Hamiltonian operator ${{{\hat{H}}}^{\left( cl \right)}}$ follows 
\begin{align}
  {{\left[ {{{\hat{w}}}^{\left( cl \right)}},{{{\hat{H}}}^{\left( cl \right)}} \right]}_{QPB}}\Psi &={{b}_{c}}{{\left[ \hat{Q},-\frac{{{\hbar }^{2}}}{2m}\frac{{{d}^{2}}}{d{{x}^{2}}}+V\left( x \right) \right]}_{QPB}}\Psi  \notag\\
 & =-\frac{{{\hbar }^{2}}}{2m}{{b}_{c}}{{\left[ \hat{Q},\frac{{{d}^{2}}}{d{{x}^{2}}} \right]}_{QPB}}\Psi +{{b}_{c}}{{\left[ \hat{Q},V\left( x \right) \right]}_{QPB}}\Psi  \notag\\
 & =\frac{{{\hbar }^{2}}}{2m}{{b}_{c}}\left( 4{{u}_{xx}}{{\Psi }_{x}}+4{{u}_{x}}{{\Psi }_{xx}}+\Psi {{u}_{xxx}} \right)+2{{b}_{c}}u{{V}_{x}}\Psi  \notag
\end{align}where $\hat{Q}=2u\frac{d}{dx}+{{u}_{x}}$ is the curvature operator, and computational process in details is given by  \[{{\left[ \hat{Q},\frac{{{d}^{2}}}{d{{x}^{2}}} \right]}_{QPB}}\Psi ={{\left[ 2u\frac{d}{dx}+{{u}_{x}},\frac{{{d}^{2}}}{d{{x}^{2}}} \right]}_{QPB}}\Psi =-4{{u}_{xx}}{{\Psi }_{x}}-4{{u}_{x}}{{\Psi }_{xx}}-\Psi {{u}_{xxx}}\]
As a consequence of above derivation, we get the time evolution of the one-dimensional G-dynamics ${{\hat{w}}^{\left( cl \right)}}$ specifically expressed as
\begin{align}
  & d{{{\hat{w}}}^{\left( cl \right)}}/dt=-\frac{{{\hbar }^{2}}}{{{m}^{2}}}\left( \left( {{u}_{x}}-\frac{{{u}^{2}}}{2} \right)\frac{{{d}^{2}}}{d{{x}^{2}}}+\left( {{u}_{xx}}-2u{{u}_{x}} \right)\frac{d}{dx}+\frac{1}{4}{{u}_{xxx}}-{{u}_{x}}^{2}-u{{u}_{xx}} \right) \notag\\
 & \begin{matrix}
   {} & {} & {}& {} & {}& {} & {}& {} & {}  \\
\end{matrix}-\frac{1}{m}\left( u{{V}_{x}}+{{u}^{2}}V \right) \notag
\end{align}Note that the time evolution of the one-dimensional G-dynamics ${{\hat{w}}^{\left( cl \right)}}$ is deeply controlled by the variable coefficients which are related to the line curvature $u$ and its derivative from the first order to three order, it shows us a complex intrinsic relation. 
Similarly, if we consider a free particle, it means $V=0$, then 
\[\frac{d}{dt}\left( {{{\hat{w}}}^{\left( cl \right)}}\Psi  \right)=-\frac{{{\hbar }^{2}}}{{{m}^{2}}}\left( \left( {{u}_{x}}-\frac{{{u}^{2}}}{2} \right)\frac{{{d}^{2}}}{d{{x}^{2}}}+\left( {{u}_{xx}}-2u{{u}_{x}} \right)\frac{d}{dx}+\frac{1}{4}{{u}_{xxx}}-{{u}_{x}}^{2}-u{{u}_{xx}} \right)\Psi \]
and 
\[d{{\hat{w}}^{\left( cl \right)}}/dt=-\frac{{{\hbar }^{2}}}{{{m}^{2}}}\left( \left( {{u}_{x}}-\frac{{{u}^{2}}}{2} \right)\frac{{{d}^{2}}}{d{{x}^{2}}}+\left( {{u}_{xx}}-2u{{u}_{x}} \right)\frac{d}{dx}+\frac{1}{4}{{u}_{xxx}}-{{u}_{x}}^{2}-u{{u}_{xx}} \right)\]

Hence, if we consider a state of equilibrium $\frac{d}{dt}\left( {{{\hat{w}}}^{\left( cl \right)}}\Psi  \right)=0$, then the second order linear nonhomogeneous differential equations with variable coefficients appears
\[\left( {{u}_{x}}-\frac{{{u}^{2}}}{2} \right){{\Psi }_{xx}}+\left( {{u}_{xx}}-2u{{u}_{x}} \right){{\Psi }_{x}}+\left( \frac{1}{4}{{u}_{xxx}}-{{u}_{x}}^{2}-u{{u}_{xx}} \right)\Psi +\frac{mu}{{{\hbar }^{2}}}\left( {{V}_{x}}+uV \right)\Psi =0\]
And for the condition $V=0$, it becomes simply 
\[\left( {{u}_{x}}-\frac{{{u}^{2}}}{2} \right){{\Psi }_{xx}}+\left( {{u}_{xx}}-2u{{u}_{x}} \right){{\Psi }_{x}}+\left( \frac{1}{4}{{u}_{xxx}}-u{{u}_{xx}}-{{u}_{x}}^{2} \right)\Psi =0\]

In order to simplify the above time evolution of the one-dimensional G-dynamics, we   reconsider the time evolution of the one-dimensional G-dynamics ${{\hat{w}}^{\left( cl \right)}}$ with the variable coefficients in another symbol,
\[\frac{d{{{\hat{w}}}^{\left( cl \right)}}}{dt}=-\frac{{{\hbar }^{2}}}{{{m}^{2}}}\left( {{A}_{0}}\frac{{{d}^{2}}}{d{{x}^{2}}}+{{A}_{1}}\frac{d}{dx}+{{A}_{2}} \right)-\left( u{{V}_{x}}+{{u}^{2}}V \right)/m\]
where we have used 
\begin{align}
  & {{A}_{0}}={{u}_{x}}-{u}^{2}/2 \notag\\
 & {{A}_{1}}={{u}_{xx}}-2u{{u}_{x}} \notag\\
 & {{A}_{2}}=\frac{1}{4}{{u}_{xxx}}-u{{u}_{xx}}-{{u}_{x}}^{2} \notag
\end{align}In general, the equation of the second order variable coefficient
satisfies ${{A}_{0}}\neq 0$, namely, ${{u}_{x}}\ne \frac{{{u}^{2}}}{2}$. Conversely, if ${{u}_{x}}=\frac{{{u}^{2}}}{2}$, then $u=-\frac{2}{x-{{C}_{12}}}$, ${C}_{12}$ is a constant, therefore, ${{A}_{0}}\neq 0$ means $u\neq-\frac{2}{x-{{C}_{12}}}$. Meanwhile, ${{A}_{0}}=0$ can result in 
\[{{A}_{0}}=0,~{{A}_{1}}={{u}_{xx}}-{{u}^{3}},~{{A}_{2}}=\frac{1}{4}{{u}_{xxx}}-\frac{1}{4}{{u}^{4}}-u{{u}_{xx}}\]
Notice that there exists a relation between the ${A}_{0}$ and ${{\hat{w}}^{\left( cl \right)}}1$ such that 
\[{{A}_{0}}=\frac{{{{\hat{w}}}^{\left( cl \right)}}1-{{w}^{\left( s \right)}}/4}{{{b}_{c}}}\]as it expresses. 
For a given function $\Psi$, it displays 
\[\frac{d}{dt}\left( {{{\hat{w}}}^{\left( cl \right)}}\Psi  \right)=-\frac{{{\hbar }^{2}}}{{{m}^{2}}}\left( {{A}_{0}}{{\Psi }_{xx}}+{{A}_{1}}{{\Psi }_{x}}+{{A}_{2}}\Psi  \right)-\frac{1}{m}\left( u{{V}_{x}}+{{u}^{2}}V \right)\Psi \]

\section{G-dynamics on coordinate transformation}
\ \ \ In this section, we start to consider how the rule of the transformation is on the coordinate transformation $y=y\left( x \right)$ of G-dynamics ${{\hat{w}}^{\left( cl \right)}}$, and we denote curvature operator \eqref{a8} as
${{\hat{Q}}_{(x)}}=2u\frac{d}{dx}+\frac{du}{dx}$ for convenient discussions, basically, derivative operator $\frac{d}{dx}={{y}_{x}}\frac{d}{dy}$ holds, and it accordingly yields $u=\frac{ds}{dx}={{s}_{x}}={{y}_{x}}\frac{ds}{dy}={{y}_{x}}{{s}_{y}}$ if we assume that $s=s\left( y\left( x \right) \right)$ is given, where ${{y}_{x}}=dy/dx$, as a result, we can obtain the deformation of the curvature operator \eqref{a8} on the coordinate transformation,
\begin{align}\label{a9}
 {{\hat{Q}}_{(x)}} & =2{{s}_{x}}\frac{d}{dx}+{{s}_{xx}}={{y}_{x}}\left( 2u\frac{d}{dy}+\frac{du}{dy} \right) \notag\\
 & ={{y}_{x}}^{2}\left( 2{{s}_{y}}\frac{d}{dy}+{{s}_{yy}} \right)+{{y}_{x}}{{s}_{y}}\frac{d{{y}_{x}}}{dy} \\
 & ={{y}_{x}}^{2}{{\hat{Q}}_{(y)}}+\frac{1}{2}{{s}_{y}}\frac{d\left( {{y}_{x}}^{2} \right)}{dy} \notag
\end{align}
where $u_{x}={{s}_{xx}}$ has been used, and ${{\hat{Q}}_{(y)}}=2{{s}_{y}}\frac{d}{dy}+{{s}_{yy}}$.    Therefore, \eqref{a9} can be written in a compact form
\[{{\hat{Q}}_{(x)}}={{y}_{x}}^{2}\left( {{\hat{Q}}_{(y)}}+\frac{1}{2}{{s}_{y}}\frac{d\left( \ln {{y}_{x}}^{2} \right)}{dy} \right)\]for all ${{y}_{x}}\ne 0$. Conversely, the ${{\hat{Q}}_{(y)}}$ is expressed as  \[{{\hat{Q}}_{(y)}}={{y}_{x}}^{-2}{{\hat{Q}}_{(x)}}-\frac{1}{2}{{s}_{y}}\frac{d\left( \ln {{y}_{x}}^{2} \right)}{dy}\]
In particular, if $\frac{d{{y}_{x}}}{dy}=0$ is given, then \eqref{a9} can be rewritten in a simple form
${{\hat{Q}}_{(x)}}={{y}_{x}}^{2}{{\hat{Q}}_{(y)}}$. Conversely, its inverse
transformation is ${{\hat{Q}}_{(y)}}=y_{x}^{-2}{{\hat{Q}}_{(x)}}$ holds for all ${y}_{x}\neq 0$.
On such results of the transformation given by \eqref{a9}, the G-dynamics \eqref{e1}  ${{\hat{w}}^{\left( cl \right)}}={{b}_{c}}\left( 2ud/dx+{{u}_{x}} \right)$ is denoted as \[\hat{w}_{\left( x \right)}^{\left( cl \right)}={{b}_{c}}{{\hat{Q}}_{\left( x \right)}}={{b}_{c}}\left( 2{{s}_{x}}\frac{d}{dx}+{{s}_{xx}} \right)\] in order to distinguish from $y$, then it becomes
\[\hat{w}_{(x)}^{\left( cl \right)}={{b}_{c}}{{\hat{Q}}_{(x)}}={{y}_{x}}^{2}\hat{w}_{(y)}^{\left( cl \right)}+\frac{{{b}_{c}}}{2}{{s}_{y}}\frac{d\left( {{y}_{x}}^{2} \right)}{dy}\]
where $\hat{w}_{(y)}^{\left( cl \right)}={{b}_{c}}{{\hat{Q}}_{(y)}}$, and
$\hat{w}_{(x)}^{\left( cl \right)}={{y}_{x}}^{2}\hat{w}_{(y)}^{\left( cl \right)}$  holds if $\frac{d{{y}_{x}}}{dy}=0$. Obviously, if identity transformation
$y=x$ is given, then ${{\hat{Q}}_{(y)}}={{\hat{Q}}_{(x)}}$. More generally, we can prove following corollary.

\begin{corollary}\label{c2}
  ${{\hat{Q}}_{(x)}}={{y}_{x}}^{2}{{\hat{Q}}_{(y)}}$ holds if and only if $y={{a}_{1}}x+{{a}_{2}}$, where ${{a}_{1}}\neq0,{{a}_{2}}$ are constants.
\begin{proof}
 Since the formula \eqref{a9} is given, hence, ${{\hat{Q}}_{(x)}}={{y}_{x}}^{2}{{\hat{Q}}_{(y)}}$ holds if and only if $d{{y}_{x}}/dy=0$, then ${{y}_{x}}={{a}_{1}}\neq0$ is a constant which has the obvious solution $y={{a}_{1}}x+{{a}_{2}}$, where ${{a}_{2}}$ is a constant.

\end{proof}
\end{corollary}
This corollary \ref{c2} says that linear transformation between $x,y$ can let relation ${{\hat{Q}}_{(x)}}={{y}_{x}}^{2}{{\hat{Q}}_{(y)}}$ hold only, in this case, we get  ${{\hat{Q}}_{(x)}}={{a}_{1}}^{2}{{\hat{Q}}_{(y)}}$.  Conversely, it reveals that all nonlinear transformation between $x,y$ make corollary \ref{c2} false.
As a matter of fact, we notice that if transformation $u\to x$ is given for the curvature operator $\hat{Q}$, then we can see that the corresponding transformation follows
$$\hat{Q}=2ud/dx+u_{x}\to \hat{\theta }/2=xd/dx+1/2$$
by this transformation, we can easily observe how unusual the G-dynamics is, in particular, it potentially reveals the hidden information of the curvature operator $\hat{Q}$ and the G-dynamics ${{\hat{w}}^{\left( cl \right)}}$.

In the below discussions, we give some examples to verify the G-dynamics ${{\hat{w}}^{\left( cl \right)}}$  on coordinate transformation \eqref{a9},

\vspace{.5 cm}
{\bf{Example 1:}} Given an exponential function $y\left( x \right)={{e}^{x}}$, $x\in \mathbb{R}$ with its inverse function $x=\ln y, y>0$,  then \eqref{a9} turns to the certain form
\[{{\hat{Q}}_{(x)}}={{y}^{2}}\left( {{\hat{Q}}_{(y)}}+\frac{1}{y}{{s}_{y}} \right)={{y}^{2}}{{\hat{Q}}_{(y)}}+y{{s}_{y}}=y\left( y{{\hat{Q}}_{(y)}}+{{s}_{y}} \right)\]and the inverse follows
${{\hat{Q}}_{(y)}}={{y}^{-2}}{{\hat{Q}}_{(x)}}-\frac{1}{y}{{s}_{y}}$.

\vspace{.5 cm}
{\bf{Example 2:}} Consider a quadratic function $y\left( x \right)=x^{2}+x$, $x\in \mathbb{R}$ with its inverse function $x=-1/2\pm \sqrt{y+1/4},~y\ge -1/4$,  then \eqref{a9} turns to the certain form
\[{{\hat{Q}}_{\left( x \right)}}={{\left( 2x+1 \right)}^{2}}{{\hat{Q}}_{\left( y \right)}}+2{{s}_{y}}, ~~x=-1/2+\sqrt{y+1/4}\]\[{{\hat{Q}}_{\left( x \right)}}={{\left( 2x+1 \right)}^{2}}{{\hat{Q}}_{\left( y \right)}}-2{{s}_{y}}, ~~x=-1/2-\sqrt{y+1/4}\]where ${{x}_{y}}=dx/dy=\pm {{y}_{x}}^{-1},~x\ne -1/2$ and the inverse follows
$${{\hat{Q}}_{\left( y \right)}}=\frac{{{\hat{Q}}_{\left( x \right)}}-2{{s}_{y}}}{{{\left( 2x+1 \right)}^{2}}},~x\ne -1/2$$ for the former expression, the latter is analogous, we omit.

\subsection{The condition of quantum geometric bracket vanishes}
In the following discussions, we consider what conditions hold if quantum geometric bracket disappears, namely, $G\left(s,\hat{f},\hat{g} \right)=0$ holds for operators $\hat{f},~\hat{g}$, then the quantum covariant Poisson bracket is degenerated to the quantum Poisson bracket, as such case, it gets an equality shown by
$\hat{f}{{\left[ s,\hat{g} \right]}_{QPB}}=\hat{g}{{\left[ s,\hat{f} \right]}_{QPB}}$. We use a specific example to explain that there exists such function $\Psi$ such that the quantum geometric bracket disappears and makes the quantum covariant Poisson bracket degenerate to the quantum Poisson bracket,  such as we choose
$\hat{f}={{\hat{p}}^{\left( cl \right)}}$ and $\hat{g}={{\hat{H}}^{\left( cl \right)}}$, then based on the condition $G\left(s,{{\hat{p}}^{\left( cl \right)}},{{\hat{H}}^{\left( cl \right)}}\right)=0$, we then have an equality given by
\begin{equation}\label{c9}
  {{\hat{p}}^{\left( cl \right)}}{{\left[ s,{{{\hat{H}}}^{\left( cl \right)}} \right]}_{QPB}}={{\hat{H}}^{\left( cl \right)}}{{\left[ s,{{{\hat{p}}}^{\left( cl \right)}} \right]}_{QPB}}
\end{equation}
More precisely, it leads to a clear result in accordance with the \eqref{c8},
${{\hat{p}}^{\left( cl \right)}}{{\hat{w}}^{\left( cl \right)}}={{\hat{H}}^{\left( cl \right)}}u$, furthermore, for a given function $\Psi$, it yields
\begin{align}
 {{{\hat{p}}}^{\left( cl \right)}}{{{\hat{w}}}^{\left( cl \right)}}\Psi & =-\sqrt{-1}\hbar \frac{d}{dx}\left( {{b}_{c}}\left( 2u{{\Psi}_{x}}+\Psi{{u}_{x}} \right) \right)=-\sqrt{-1}\hbar {{b}_{c}}\frac{d}{dx}\left( 2u{{\Psi}_{x}}+\Psi{{u}_{x}} \right) \notag\\
 & =-\sqrt{-1}\hbar {{b}_{c}}\left( 2u{{\Psi}_{xx}}+3{{u}_{x}}{{\Psi}_{x}}+\Psi{{u}_{xx}} \right) \notag
\end{align}
and
\begin{align}
 {{{\hat{H}}}^{\left( cl \right)}}\left( u\Psi  \right) & =-\frac{{{\hbar }^{2}}}{2m}\frac{{{d}^{2}}\left( u\Psi  \right)}{d{{x}^{2}}}+u\Psi V\left( x \right) \notag\\
 & =-\frac{{{\hbar }^{2}}}{2m}\left( {{u}_{xx}}\Psi +2{{u}_{x}}{{\Psi }_{x}}+u{{\Psi }_{xx}} \right)+u\Psi V\left( x \right) \notag
\end{align}
Hence, the equation ${{\hat{p}}^{\left( cl \right)}}{{\hat{w}}^{\left( cl \right)}}\Psi={{\hat{H}}^{\left( cl \right)}}u\Psi$ holds for some functions $\Psi$.
By simplifying the equality, we obtain the second order ODE in terms of the function $\Psi$ satisfying the differential equation \[u{{\Psi}_{xx}}+{{u}_{x}}{{\Psi}_{x}}+\sigma u\Psi=0\]where
$\sigma =\sigma \left( x \right)=\frac{2m}{{{\hbar }^{2}}}V\left( x \right)$ has denoted for convenience, subsequently, we get second order ODE expressed as ${{\Psi}_{xx}}+{{\left( \ln u \right)}_{x}}{{\Psi}_{x}}+\sigma \Psi=0$, we discuss the case without potential energy, that is $V=0$, then $\sigma =\sigma \left( x \right)=0$ moreover, it has  ${{\Psi}_{xx}}+{{\left( \ln u \right)}_{x}}{{\Psi}_{x}}=0$ follows, in order to get its solution, we let $\xi ={{\Psi}_{x}}$, then ${{\xi }_{x}}+{{\left( \ln u \right)}_{x}}\xi =0$ is obtained with solution $\xi ={{C}_{11}}{{u}^{-1}}$, thusly, the solution is accordingly given by
\begin{equation}\label{d3}
  \Psi\left( x \right)={{C}_{11}}\int{{{u}^{-1}}}dx
\end{equation}
where ${C}_{11}$ is a constant. In this example, we can surely say that there exists some functions such that the quantum geometric bracket vanishes. In other words, for
$\hat{f}{{\left[ s,\hat{g} \right]}_{QPB}}=\hat{g}{{\left[ s,\hat{f} \right]}_{QPB}}$, there exists functions $\wp$ such that
\[\left( \hat{f}s\hat{g}-\hat{g}s\hat{f} \right)\wp ={{\left[ \hat{f},\hat{g} \right]}_{QPB}}\left( s\wp  \right)\]holds for operators $\hat{f},~\hat{g} $.
In above specific case \eqref{c9}, the proper function $\wp ={{C}_{11}}\int{{{u}^{-1}}}dx$ is given. Conversely, it proves that the geometric scalar function $s$ can be chosen to be existed for the manifold space.

Given by \eqref{d1}, we obtain time evolution in terms of ${{\hat{p}}^{\left( cl \right)}}$, that is
$$d{{\hat{p}}^{\left( cl \right)}}/dt =-\left( {{V}_{x}}+{{\hat{H}}^{\left( cl \right)}}u \right)$$Then the covariant time evolution with respect to ${{\hat{p}}^{\left( cl \right)}}$ follows
\begin{align}
 \frac{\mathcal{D}{{{\hat{p}}}^{\left( cl \right)}}}{dt}\Psi & =\frac{d{{{\hat{p}}}^{\left( cl \right)}}}{dt}\Psi +{{{\hat{p}}}^{\left( cl \right)}}{{{\hat{w}}}^{\left( cl \right)}}\Psi =-\left( {{V}_{x}}+{{{\hat{H}}}^{\left( cl \right)}}u \right)\Psi -\sqrt{-1}\hbar {{b}_{c}}\left( 2u{{\Psi }_{xx}}+3{{u}_{x}}{{\Psi }_{x}}+\Psi {{u}_{xx}} \right) \notag\\
 & =-\left( {{V}_{x}}\Psi +{{{\hat{H}}}^{\left( cl \right)}}u\Psi  \right)-\sqrt{-1}\hbar {{b}_{c}}\left( 2u{{\Psi }_{xx}}+3{{u}_{x}}{{\Psi }_{x}}+\Psi {{u}_{xx}} \right) \notag\\
 & =-{{V}_{x}}\Psi -u\Psi V\left( x \right)+\frac{{{\hbar }^{2}}}{2m}\left( {{u}_{xx}}\Psi +2{{u}_{x}}{{\Psi }_{x}}+u{{\Psi }_{xx}}-2u{{\Psi }_{xx}}-3{{u}_{x}}{{\Psi }_{x}}-\Psi {{u}_{xx}} \right) \notag\\
 & =-{{V}_{x}}\Psi -u\Psi V\left( x \right)-\frac{{{\hbar }^{2}}}{2m}\left( u{{\Psi }_{xx}}+{{u}_{x}}{{\Psi }_{x}} \right) \notag
\end{align}
Consider equation of the covariant equilibrium $\frac{\mathcal{D}{{{\hat{p}}}^{\left( cl \right)}}}{dt}\Psi =0$, then it yields the second order ODE that is satisfying the equation
\begin{equation}\label{d4}
  {{\Psi }_{xx}}+\frac{{{u}_{x}}}{u}{{\Psi }_{x}}+\chi \Psi =0
\end{equation}
where
$\chi =\chi \left( x \right)=\frac{2m}{{{\hbar }^{2}}}\left( {{V}_{x}}/u+V\left( x \right) \right)$ has been denoted. If we assume that $V=0$, then $\chi =0$ and \eqref{d4} can be simplified as \[{{\Psi }_{xx}}+\frac{{{u}_{x}}}{u}{{\Psi }_{x}} =0\]
Its solution is previously obtained and given by \eqref{d3},  as described by  \eqref{d3} and  \eqref{d4} in different conditions, we can observe that the different part is about the potential energy $V$ and its first derivative $V_{x}$ in terms of the coordinates.

\section{Conclusions}
In conclusions, by giving the certain results on the analysis of two different kinds of one-dimensional G-dynamics ${{\hat{w}}^{\left( cl \right)}}={{b}_{c}}\left( 2ud/dx+K \right)$  and the curvature operator $\hat{Q}=2ud/dx+K$, and ${{\hat{w}}^{\left( cl \right)}}1={b}_{c}K$, and another G-dynamics is ${{\hat{w}}^{\left( ri \right)}}={{b}_{c}}\left( 2ud/dx+K+2{{u}^{2}} \right)$. Therefore, as ${{{\hat{w}}}^{\left( cl \right)}}\to {{{\hat{w}}}^{\left( ri \right)}}={{{\hat{w}}}^{\left( cl \right)}}+{{w}^{\left( s \right)}}$ shows, the Hermitian Hamiltonian operators  ${{{\hat{H}}}^{\left( cl \right)}}$ correspondingly turns to non-Hermitian Hamiltonian operators ${{{\hat{H}}}^{\left( ri \right)}}$, the relation reads ${{{\hat{H}}}^{\left( cl \right)}}\to {{{\hat{H}}}^{\left( ri \right)}}={{{\hat{H}}}^{\left( cl \right)}}-{{E}^{\left( s \right)}}/2-\sqrt{-1}\hbar {{{\hat{w}}}^{\left( cl \right)}}$.    As proven, it is turned out that there are so many good properties of the G-dynamics, and the similar mode of the G-dynamics \eqref{e1} are widely used such as \eqref{c3} $\Delta \hat{a}$ and \eqref{c4} $\hat{\alpha }$ as mentioned previously. The G-dynamics on coordinate transformation shows us a nonlinear relation between two coordinates system. It is interesting to note that similarities do exist between the Schr\"{o}dinger equation and the geometric wave equation.  Among the defining characteristics about the G-dynamics \eqref{e1}, this operator works almost in complete different way.
Moreover, there are still having more hidden properties left to be found about the
G-dynamics \eqref{e1}. As a first order differential operator, there will be more further consequences, such as the exactly meanings of the second order variable coefficients from the time evolution of the G-dynamics ${{\hat{w}}^{\left( cl \right)}}$,
\begin{align}
  & {{A}_{0}}={{u}_{x}}-\frac{{{u}^{2}}}{2},~~{{A}_{1}}={{u}_{xx}}-2u{{u}_{x}} ,~~ {{A}_{2}}=\frac{1}{4}{{u}_{xxx}}-u{{u}_{xx}}-{{u}_{x}}^{2} \notag
\end{align}with relation ${{A}_{0}}=\left( {{{\hat{w}}}^{\left( cl \right)}}1-{{w}^{\left( s \right)}}/4 \right)/{{b}_{c}}$.    
As with many other operators, the G-dynamics \eqref{e1} has its own characters, it's existed, which serves as something of a physical front for the quantum mechancis, in periodical form.

\section*{References}

\ \ \ \ [DG] Griffiths, J David. Introduction to Quantum Mechanics (2nd ed.). Prentice Hall, 2004.\par
[GW] Gen Wang. Generalized geometric commutator theory and quantum geometric bracket and its uses. \par arXiv:2001.08566. \par
[SW] Stephen Wiggins. Introduction to Applied Nonlinear Dynamical Systems and Chaos, (2nd ed). Springer Verlag,  2003.\par
[TF] Theodore Frankel. The Geometry of Physics-An introduction.
Cambridge University Press, 2004.

\end{document}